\documentclass[prc,showpacs]{revtex4}% Physical Review C
\usepackage{graphicx}% Include figure files
\usepackage{dcolumn}% Align table columns on decimal point
\usepackage{bm}% bold math
\newcommand {\car}{$^{12}$C }
\newcommand {\oxy}{$^{16}$O }
\newcommand {\caI}{$^{40}$Ca }
\newcommand {\caII}{$^{48}$Ca }
\newcommand {\lead}{$^{208}$Pb }
\newcommand{\br}{{\bf r}}
\newcommand{\half} {\frac{1}{2}}
\newcommand{\bsigma}{\mbox{\boldmath $\sigma$}}
\newcommand{\btau}{\mbox{\boldmath $\tau$}}

\newcommand{\ton} {t_1}
\newcommand{\ttw} {t_2}

\begin{document}
\title{Momentum distributions and spectroscopic factors 
of doubly-closed shell nuclei in correlated basis function theory}
\author{C. Bisconti$^{\,1,2,3}$, 
F. Arias de Saavedra$^{\,2}$ and G. Co'$^{\,1}$}
\affiliation{
$^1$ Dipartimento di Fisica, Universit\`a del Salento 
 and I.N.F.N., sezione di Lecce, 
I-73100 Lecce, Italy \\
$^2$ Departamento de F\'{\i}sica
At\'omica, Molecular y Nuclear, Universidad de Granada, 
E-18071 Granada, Spain \\
$^3$ Dipartimento di Fisica, Universit\`a di Pisa, 
I-56100 Pisa, Italy 
}

\date{\today}

\begin{abstract}
  The momentum distributions, natural orbits, spectroscopic factors
  and quasi-hole wave functions of the \car,\oxy,\caI,\caII, and \lead
  doubly closed shell nuclei, have been calculated in the framework of
  the Correlated Basis Function theory, by using the Fermi hypernetted
  chain resummation techniques.  The calculations have been done by
  using the realistic Argonne $v'_8$ nucleon-nucleon potential,
  together with the Urbana IX three-body interaction. Operator
  dependent correlations, which consider channels up to the tensor
  ones, have been used.  We found noticeable effects produced by the
  correlations. For high momentum values, the momentum distributions
  show large enhancements with respect to the independent particle
  model results. Natural orbits occupation numbers are depleted by
  about the 10\% with respect to the independent particle model
  values.  The effects of the correlations on the spectroscopic
  factors are larger on the more deeply bound states.
\end{abstract}

\maketitle

\section{INTRODUCTION}
\label{sec:int}
One of the major achievements of nuclear structure studies in the last
ten years is the consolidation of the validity of the non relativistic
many-body approach. The idea is to describe the nucleus with a
Hamiltonian of the type:
\begin{equation}
H=-\frac{\hbar^2}{2m}\sum_{i}^A\nabla_i^2+ 
\sum_{i<j=1}^A v_{ij}+\sum_{i<j<k=1}^A v_{ijk}
\,\,,
\label{eq:hamiltonian}
\end{equation}
where the two- and three-body interactions, $v_{ij}$ and $v_{ijk}$
respectively, are fixed to reproduce the properties of the two- and
three-body nuclear systems. The Schr\"odinger equation has been solved
without approximations for few body systems \cite{kam01} and light
nuclei \cite{pud97}, up to A=12 \cite{pie05}. The obtained results
provide good descriptions, not only of the energies of these nuclei,
but also of other observables.

The difficulties in extending to medium and heavy nuclei the
techniques used in few body systems and light nuclei, favored the
development of models, and of effective theories.  The basic idea of
the effective theories is to work in a restricted space of the
many-body wave functions. Usually, one works with many-body wave
functions which are Slater determinants of single particle wave
functions.  The idea of single particle wave functions implies the
hypothesis of a mean-field where each nucleon move
independently from the other ones. This Independent Particle Model (IPM)
is quite far from the picture outlined by the microscopic calculations
quoted above, describing the nucleus as a many-body system of strongly
interacting nucleons. In the Hartree-Fock theory, which provides the
microscopic foundation of the IPM, the Hamiltonian is not any more
that of Eq.  (\ref{eq:hamiltonian}), but it is an effective
Hamiltonian built to take into account, obviously in an effective
manner, the many-body effects that the microscopic calculations
explicitly consider.  The construction of effective interactions
starting from the microscopic ones, covers a wide page of the nuclear
physics history, starting from the Brueckner G-matrix effective
interactions \cite{bru55,nak86}, up to the recent $V_{low k}$
interaction \cite{bog03,cor06} and the interaction obtained by
applying the unitary correlation operator method \cite{rot04,rot06}.

The application of the IPM is quite successful, but there are
evidences of the intrinsic limitations in its applicability. For
example, the measured spectroscopic factors are systematically smaller
than one \cite{lap93,kra01,van01t}, which is the value predicted by
the IPM.  The (e,e'p) cross sections in the quasi-elastic region need
a consistent reduction of the IPM hole strength to be explained
\cite{qui88t,bof96}.  The same holds for the electromagnetic form
factors of the low-lying states, especially those having large angular
momentum \cite{lic79,hyd87}.  The emission of two like nucleons in
photon and electron scattering process cannot be described by the IPM
\cite{ond97,ond98}.  Also the charge density distributions extracted
by elastic electron scattering data are, in the nuclear interior,
smaller than those predicted by the IPM \cite{cav82,pap86}.  These
examples indicate the presence of physics effects, commonly called
correlations, which are not described by the IPM.

It is common practice to distinguish between long- and short-range
correlations since they have different physical sources. The long-range
correlations are related to collective excitations of the system, such
as the giant resonances. The short-range correlations (SRC) are
instead connected to the strongly repulsive core of the microscopic
nucleon-nucleon interaction. The repulsive core reduces the
possibility that two nucleons can approach each other, and this
modifies the IPM picture where, by definition, the motion of each
nucleon does not depend on the presence of the other ones.

Even though most of the calculations of medium heavy nuclei
are based on the IPM, and on the effective theories, various  
techniques, aiming to attack the problem by using the microscopic
Hamiltonian (\ref{eq:hamiltonian}), have been developed.  The
Brueckner-Hartree-Fock approach has been recently applied to the
\oxy nucleus \cite{dic04}.  No core-shell model calculations have been
done for nuclei lighter than \car \cite{nav00,for05}.  The coupled
cluster method has been used to evaluate \oxy properties
\cite{hei99,mih99}.

About fifteen years ago \cite{co92}, we started a project aimed to
apply to the description of medium and heavy nuclei the Correlated
Basis Function (CBF) theory, successfully used to describe the nuclear
and neutron matter properties \cite{wir88,akm98}.  We solve the
many-body Schr\"odinger equation by using the variational principle:
\begin{equation}
\delta E[\Psi]=
\delta \frac{<\Psi|H|\Psi>}{<\Psi|\Psi>} = 0 \,\,.
\label{eq:varprin}
\end{equation}
The search for the minimum of the energy functional is done within a
subspace of the full Hilbert space spanned by the A-body wave
functions which can be expressed as:
\begin{equation}
\Psi(A)={\cal F}(1,...,A)\Phi(1,...,A) \,\,,
\label{eq:psi}
\end{equation}
where ${\cal F}(1,...,A)$ is a many-body correlation operator and
$\Phi(1,...,A)$ is a Slater determinant composed by single
particle wave functions, $\phi_{\alpha}(i)$. In our calculations, we
used two-body interactions of Argonne and Urbana type, and
we considered all the interaction channels up to the spin-orbit ones.
Together with these two-body interactions, we used the appropriated
three-body forces of Urbana type.
The complexity of the interaction required the use of
operator dependent correlations. We consider correlations of the
type: 
\begin{equation}
{\cal F}={\cal S}\left( \prod_{i<j=1}^{A}F_{ij} \right) \,\,,
\label{eq:cor1}
\end{equation}
where ${\cal S}$ is a symmetrizer operator and $F_{ij}$ is expressed
in terms of two-body correlation functions $f_p$ as:
\begin{equation}
F_{ij}=\sum_{p=1}^6 f_p(r_{ij})O^p_{ij} \,\,.
\label{eq:cor2}
\end{equation}
In the above equation we have adopted the nomenclature commonly used
in this field, by defining the operators as:
\begin{equation}
O^{p=1,6}_{ij}=[1,\bsigma_i\cdot\bsigma_j,S_{ij}]\otimes
[1,\btau_i\cdot\btau_j] \,\,,
\end{equation}
where $\bsigma_i$ and $\btau_i$ indicate the usual Pauli spin and
isospin operators, and 
 $S_{ij}=(3\hat{\br}_{ij}\cdot\bsigma_i\hat{\br}_{ij}\cdot\bsigma_j
-\bsigma_i\cdot\bsigma_j)$ is the tensor operator.

We recently succeeded in formulating the Fermi Hypernetted Chain
(FHNC) equations, in Single Operator Chain (SOC) approximation, for
nuclei non saturated in isospin, and with single particle basis
described in a $jj$ coupling scheme. We presented in Ref.
\cite{bis06} the binding energies and the charge distributions of
\car, \oxy, \caI, \caII and \lead doubly closed shell nuclei obtained
by using the minimization procedure (\ref{eq:varprin}).  These
calculations have the same accuracy of the best variational
calculations done in nuclear and neutron matter \cite{wir88,akm98}.

In the present article, we show the results of our study, done in the
FHNC/SOC computational scheme, on some ground state quantities related
to observables. They are momentum distributions, natural orbits and
their occupation numbers, quasi-hole wave functions and spectroscopic
factors. We used the many-body wave functions obtained in Ref.
\cite{bis06} by solving Eq. (\ref{eq:varprin}) with the Argonne
$v'_{8}$ two-nucleon potential, together with the Urbana IX three-body
force.  We have calculated momentum distributions also with the wave
functions produced by another interaction, the Urbana $v_{14}$
truncated up to the spin-orbit terms, implemented with the Urbana VII
three-body force. The results obtained with this last interaction do
not show relevant differences with those obtained with the $v'_{8}$
and UIX interaction, therefore we do not present them.

The paper is organized as follows.  In Sect. \ref{sec:obdm} we present
the results of the One-Body Density Matrix (OBDM) and of the momentum
distribution.  In Sect. \ref{sec:no} we discuss the natural orbits,
i.e.  the single particle wave functions forming the basis where the
OBDM is diagonal.  In Sect.  \ref{sec:specf} we present our results
about the quasi-hole wave functions and in Sect.  \ref{sec:summary} we
summarize our results and draw our conclusions.

\section{ONE-BODY DENSITY MATRIX AND MOMENTUM DISTRIBUTION} 
\label{sec:obdm}
We define the one-body density matrix, (OBDM), of a system of $A$
nucleons as:
\begin{eqnarray}
\nonumber
\rho(x_1,x'_{1})  & \equiv & 
\sum_{s,s',t} \rho^{s,s';t}(\br_1,\br'_{1}) \, 
\chi^{\dagger}_s(1) \chi^{\dagger}_t(1)
\chi_{s'}(1') \chi_t(1') \\
& \equiv & \frac{A}{<\Psi|\Psi>} 
\int dx_2  \ldots dx_A 
\Psi^{\dagger}(x_1,x_2,\ldots,x_A) 
 \Psi(x'_1,x_2,\ldots,x_A) \,\,\,.
\label{eq:obdm}
\end{eqnarray}
In the above expression, the variable $x_i$ indicates the position
($\br_i$) and the third components of the spin ($s_i$) and of the
isospin ($t_i$) of the single nucleon. The $\chi(i)$ functions
represent the Pauli spinors.  With the integral sign we understand
that also the sum on spin and isospin third components of all the
particles from 2 up to A, is done.  In our calculations we are
interested in the quantity:
\begin{equation}
\rho^t(\br_1,\br'_{1}) =
\sum_{s=\pm 1/2} \left[
\rho^{s,s;t}(\br_1,\br'_{1}) + \rho^{s,-s;t}(\br_1,\br'_{1})
\right] \,\,,
\label{eq:obdm1}
\end{equation}
whose diagonal part ($\br'_1=\br_1$) represents the one-body density of 
neutrons or protons.

We obtain the momentum distributions of protons ($t$=1/2) or neutrons
($t$=-1/2) as:
\begin{equation}
n^t(k)= \frac {1}{(2 \pi)^3}
\frac {1} {{\cal N}_t} \int d\br_1 d\br'_1 \, 
e^{i {\bf k} \cdot (\br_1-\br'_1)} 
\rho^t(\br_1,\br'_1) \,\,,
\label{eq:md}
\end{equation}
where we have indicated with ${\cal N}_t$  the number of protons 
or neutrons. The above definitions imply the following
normalization of $n(k)$:
\begin{equation}
\int d{\bf k} \,n^t(k)= 1  \,\,.
\label{eq:nor}
\end{equation}

We describe doubly closed shell nuclei, with different numbers of
proton and neutrons, in a $jj$ coupling scheme. The most efficient
single particle basis to be used is constructed by a set of single
particle wave functions expressed as:
\begin{equation}
\phi^t_{nljm}(\br_i) = R^t_{nlj}(r_i) \sum_{\mu,s} <l \mu \half s| j m> 
Y_{l\mu}(\Omega_i) \chi_s(i)\chi_t(i)= R^t_{nlj}(r_i) 
{\bf Y}^m_{lj} (\Omega_i)\chi_t(i) \,\,.
\label{eq:spwf}
\end{equation}
In the above expression we have indicated with $Y_{l\mu}$ the
spherical harmonics, with $< | >$ the Clebsch-Gordan coefficient, with
$R^t_{nlj}(r_i)$ the radial part of the wave function, and with ${\bf
  Y}^m_{lj}$ the spin spherical harmonics \cite{edm57}.

The uncorrelated OBDMs, those of the IPM, are obtained by substituting in
Eq. (\ref{eq:obdm}) the correlated function $|\Psi>$ with 
the Slater determinant $|\Phi>$ formed by the
single particle wave functions (\ref{eq:spwf}). We obtain the
following expressions:
\begin{equation}
\nonumber
\rho_o^t(\br_1,\br'_1) = \sum_{s} \left[ \rho_o^{s,s;t}(\br_1,\br'_1)
+\rho_{o}^{s,-s;t}(\br_1,\br'_1) \right]  \,\,,
\label{eq:obdm0}
\end{equation}
\begin{eqnarray}
\rho_o^{s,s;t}(\br_1,\br'_1) 
&=&
               \frac{1}{8\pi}\sum_{nlj}(2j+1)R^t_{nlj}(r_1)
                R^t_{nlj}(r'_1)P_{l}(\cos \theta_{11'}) \,\,,
\label{eq:obdmp}
\\
\rho_{o}^{s,-s;t}(\br_1,\br'_1) 
&=&\frac{1}{4\pi}
\sum_{nlj}(-1)^{j-l-1/2}R^t_{nlj}(r_1)R^t_{nlj}(r'_1)
                \sin\theta_{11'}P'_{l}(\cos \theta_{11'}) \,\,.
\label{eq:obdma}
\end{eqnarray}
In the above equations $\theta_{11'}$ indicates the angle between
$\br_1$ and $\br'_1$, and $P_l$ and $P'_l$ the Legendre polynomials
and their first derivative respectively.  The presence of the second
term of Eq. (\ref{eq:obdm0}), the antiparallel spin terms given in Eq.
(\ref{eq:obdma}), is due the $jj$ coupling scheme required to describe
heavy nuclei.

\begin{figure}[ht]
\includegraphics [scale=0.60]
                 {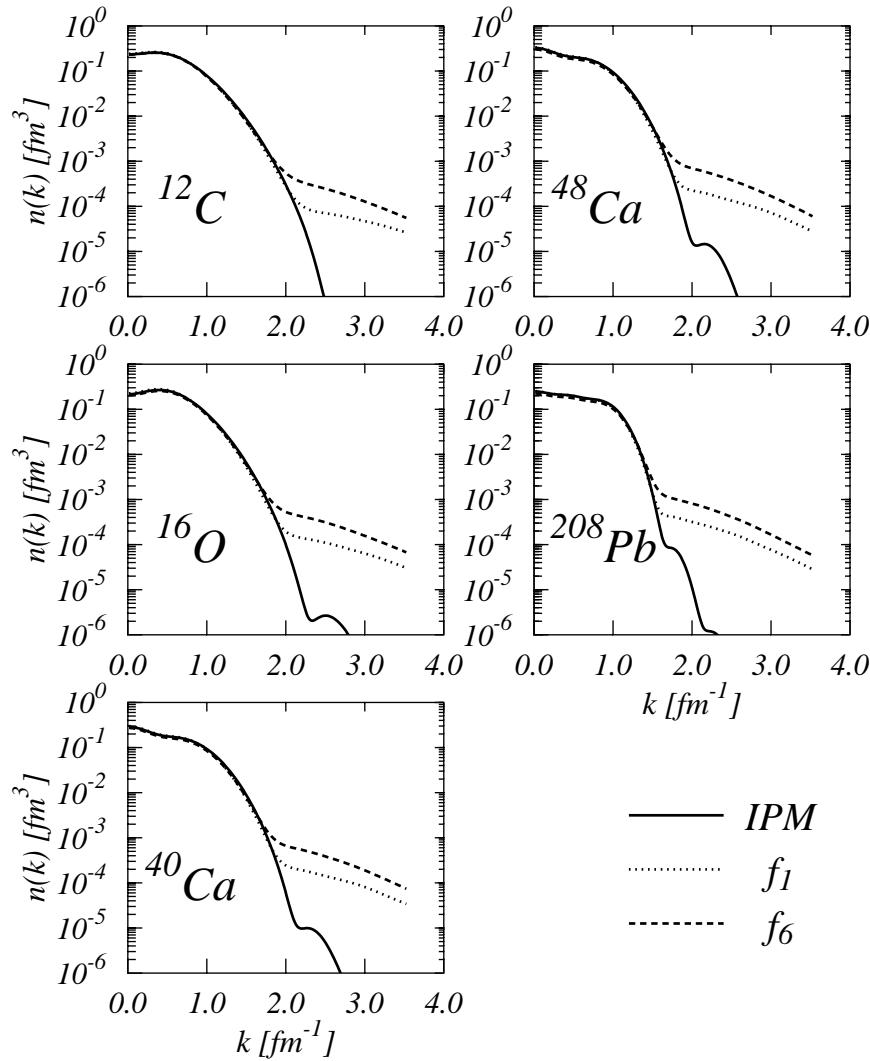}
\caption{\small The proton momentum distributions 
  of the \car, \oxy, \caI, \caII and \lead nuclei calculated in the
  IPM model, by using the scalar correlations only
  ($f_1$) and the full operator dependent correlations 
  ($f_6$).  
}
\label{fig:md} 
\end{figure}

The correlated OBDM is obtained by using the ansatz (\ref{eq:psi}) in
Eq. (\ref{eq:obdm}). This calculation is done by using the cluster
expansion techniques as indicated in \cite{co94} and \cite{ari96},
where only scalar correlations have been used, and in \cite{fab01},
where the state dependent correlations have been used, but in a $ls$
coupling scheme. We followed the lines of Ref. \cite{fab01} and, in
addition, we consider the presence of the antiparallel spin terms and
we distinguish proton and neutron contributions.  The explicit
expression of the OBDM, in terms of FHNC/SOC quantities, such as
two-body density distributions, vertex corrections, nodal diagrams
etc., is given in Appendix \ref{app:A}.  The diagonal part of the OBDM
is the one-body density, normalized to the number of nucleons. Because
of this, the momentum distribution satisfies the following sum rule:
\begin{equation}
S^t_2 =  \frac {\hbar^2} {2 m} 
\int d {\bf k} \, k^2 \, n^t(k) / T_{FHNC}^t = 1 \,\,,
\label{eq:sr2}
\end{equation}
where we have indicated with $T_{FHNC}^t$ the kinetic energy of the
protons or of the neutrons.  We have verified the accuracy of our
calculations by testing the normalization (\ref{eq:nor}) and the
exhaustion of the above sum rule for every $n^t(k)$ calculated. We
found that these quantities are always satisfied at the level of few
parts on a thousand in a full FHNC/SOC calculation, and even better
when only scalar correlations are used.
%
% obdm
%
\begin{figure}[pt]
\includegraphics [scale=0.52,angle=270] {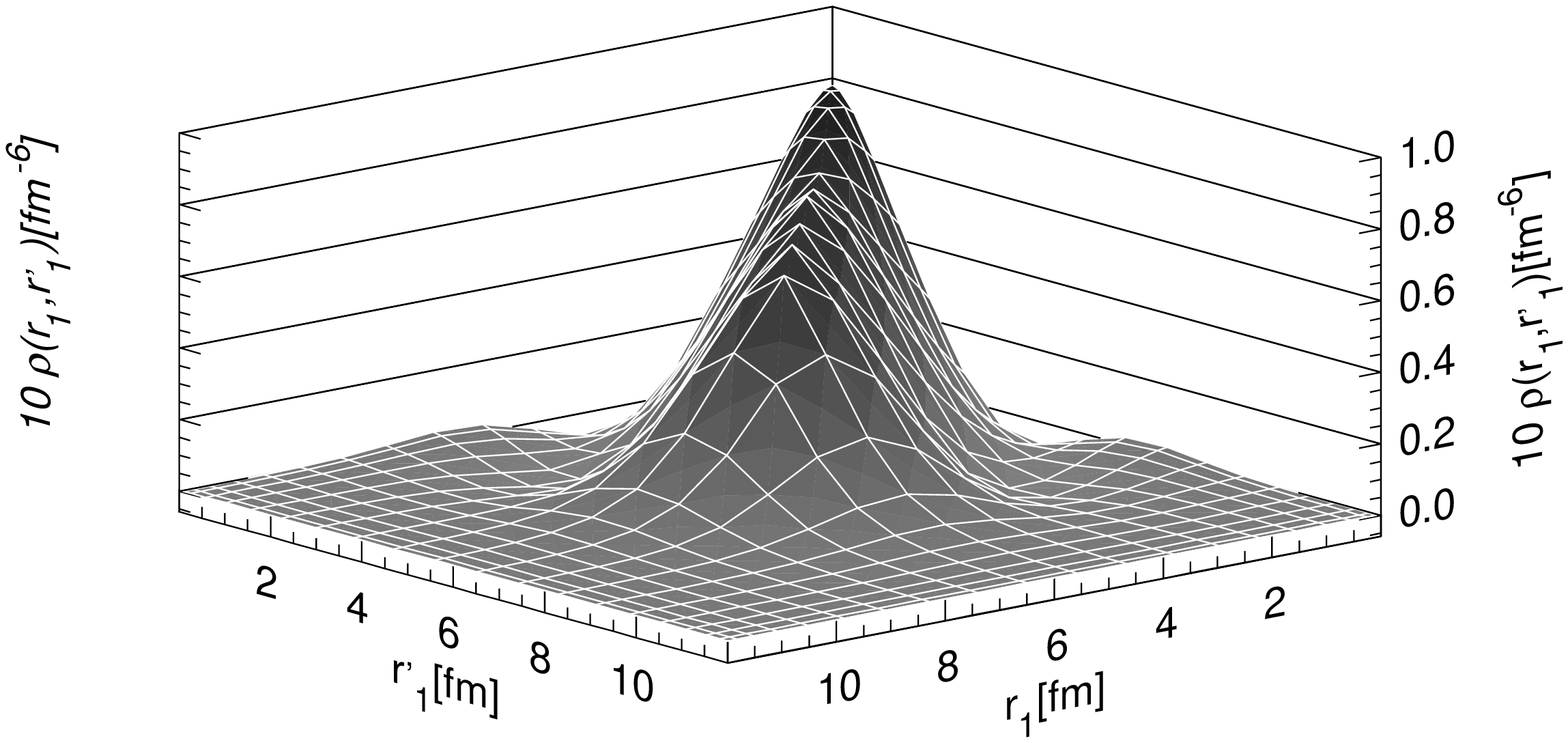}
\caption{\small The proton one-body density matrix, $\rho(r_1,r'_{1})$, 
for the \lead nucleus in FHNC/SOC approximation, 
calculated for $\theta_{11'}$=0. 
The diagonal part $\rho(r_1,r_1)$ is the proton density distribution.
}
\label{fig:obdm} 
%
% difference in obdm
%
%\end{figure}
%\begin{figure}[pt]
\includegraphics [scale=0.52,angle=270] {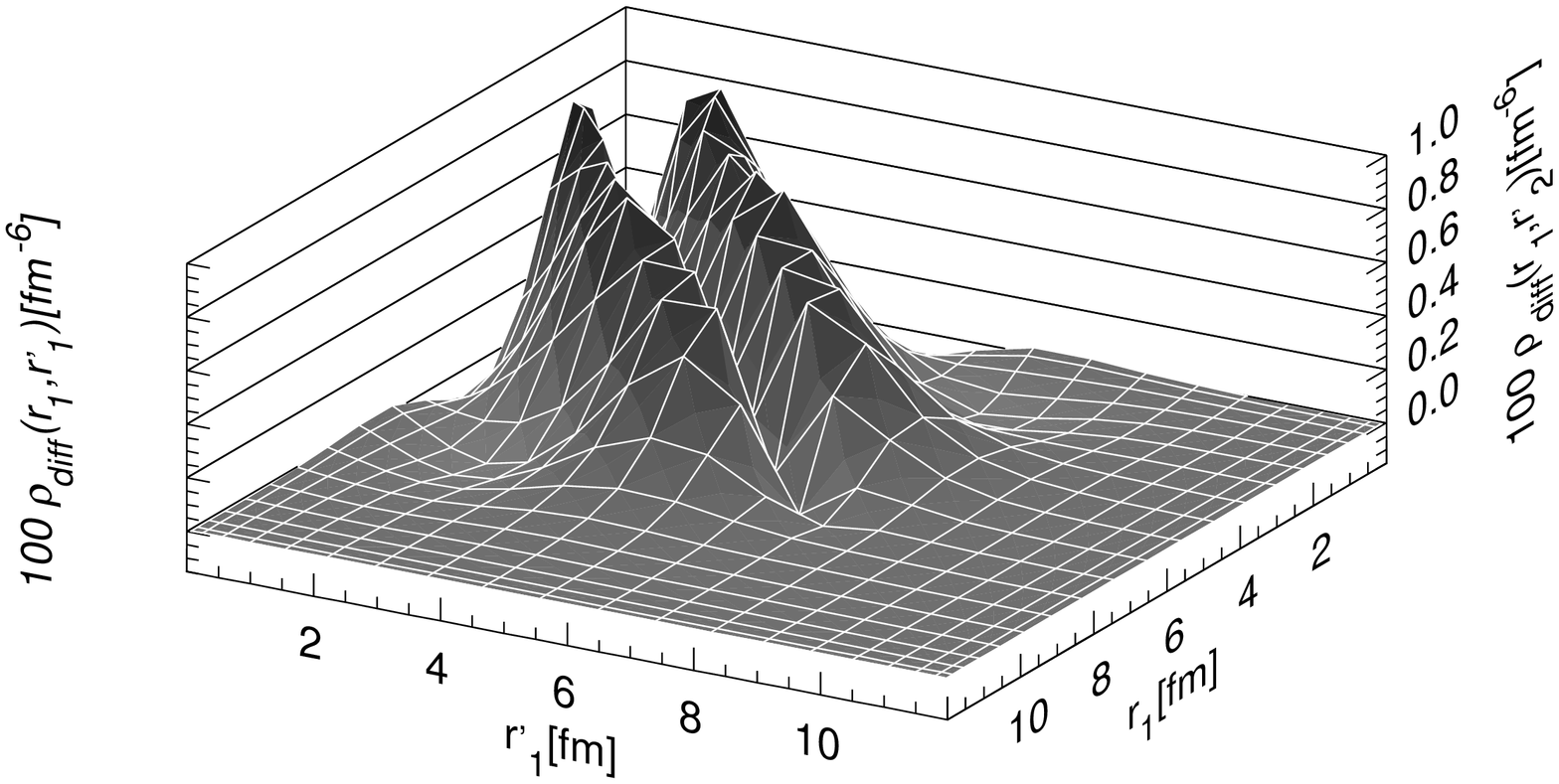}
\caption{\small The difference
$\rho_{o}(r_1,r'_1)-\rho(r_1,r'_1)$, between 
the proton IPM
one-body density matrix of the \lead nucleus,
and that obtained with our FHNC/SOC calculations. The two density
matrices, have been calculated for  $\theta_{11'}$=0. Note that the
scale here is ten times larger than that of Fig. 
\protect\ref{fig:obdm}.
}
\label{fig:obdmdiff} 
\end{figure}
The surface shown in Fig.\ref{fig:obdm} represents the proton OBDM of
the \lead nucleus, for $\theta_{11'}$=0.  We have shown in
\cite{bis06} that the SRC lower the one-body proton distribution, and
also that of the neutrons, in the nuclear center. In order to
highlight the effects of the correlations on the density matrix, we
show in Fig.  \ref{fig:obdmdiff} the quantity
$\rho_{o}(r_1,r'_1)-\rho(r_1,r'_1)$.  Note that the $z$-axis scale of
Fig. \ref{fig:obdmdiff} is ten times larger than that of
Fig.\ref{fig:obdm}.  It is interesting to notice that the major
differences between the OBDMs are not in the diagonal part, but just
beside it.  The consequences of these, small, differences between the
OBDMs on the momentum distributions, are shown in Fig. \ref{fig:md}.
In this figure, we compare the \car, \oxy, \caI, \caII and \lead
proton momentum distributions calculated in the IPM model, with those
obtained by using scalar and operator dependent correlations.
%
% momentum distributions
%
\begin{figure}[ht]
\includegraphics [scale=0.50]
                 {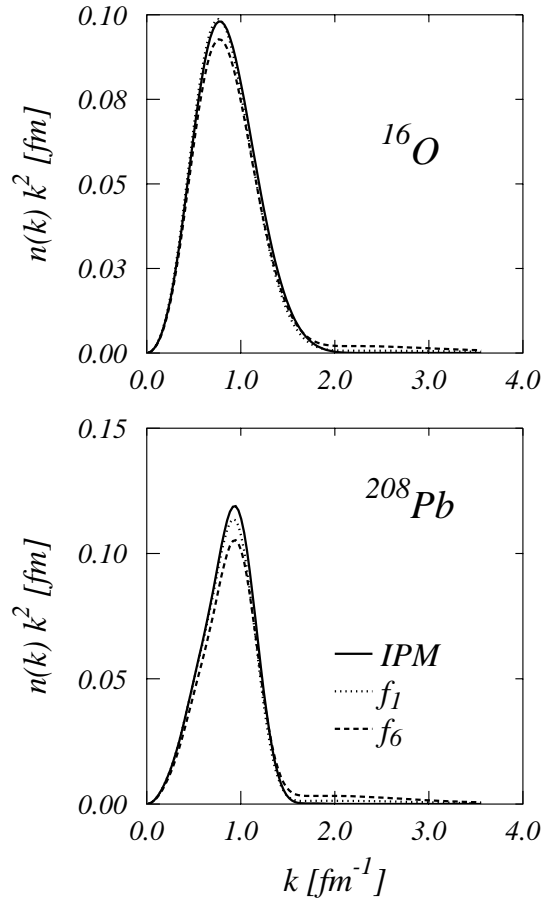}
\caption{\small The proton momentum distributions of the 
 \oxy and \lead nuclei multiplied by $k^2$. The full lines show the
 IPM results, the dotted lines have been obtained by using 
 scalar correlations only, and the dashed lines with the complete
 correlation. 
}
\label{fig:mdlin} 
\end{figure}
\begin{figure}[ht]
\includegraphics [scale=0.60]
                 {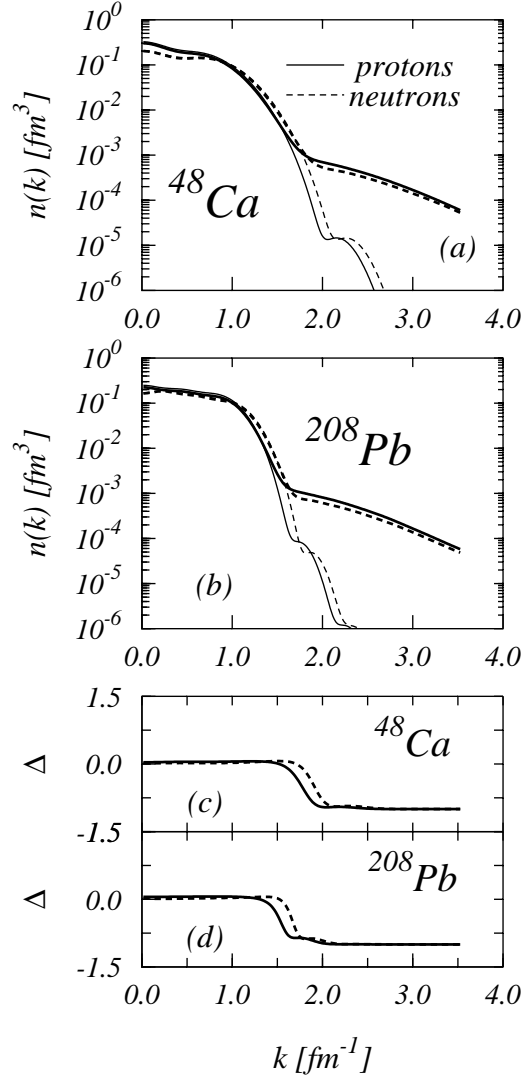}
\caption{\small In the panels (a) and (b) we show the 
  protons (full lines) and neutrons (dashed lines) momentum
  distributions of the \caII and \lead. The thick lines show the
  results of our calculations, the thin lines the IPM results.  In the
  panels (c) and (d), we show the weighted difference
  (\protect\ref{eq:delta}) between uncorrelated and correlated
  momentum distributions. As in the upper part of the figure, the full
  lines show the protons results and the dashed lines those of the
  neutrons.  }
\label{fig:mdpn} 
\end{figure}
\begin{figure}[ht]
\includegraphics [scale=0.50]
                 {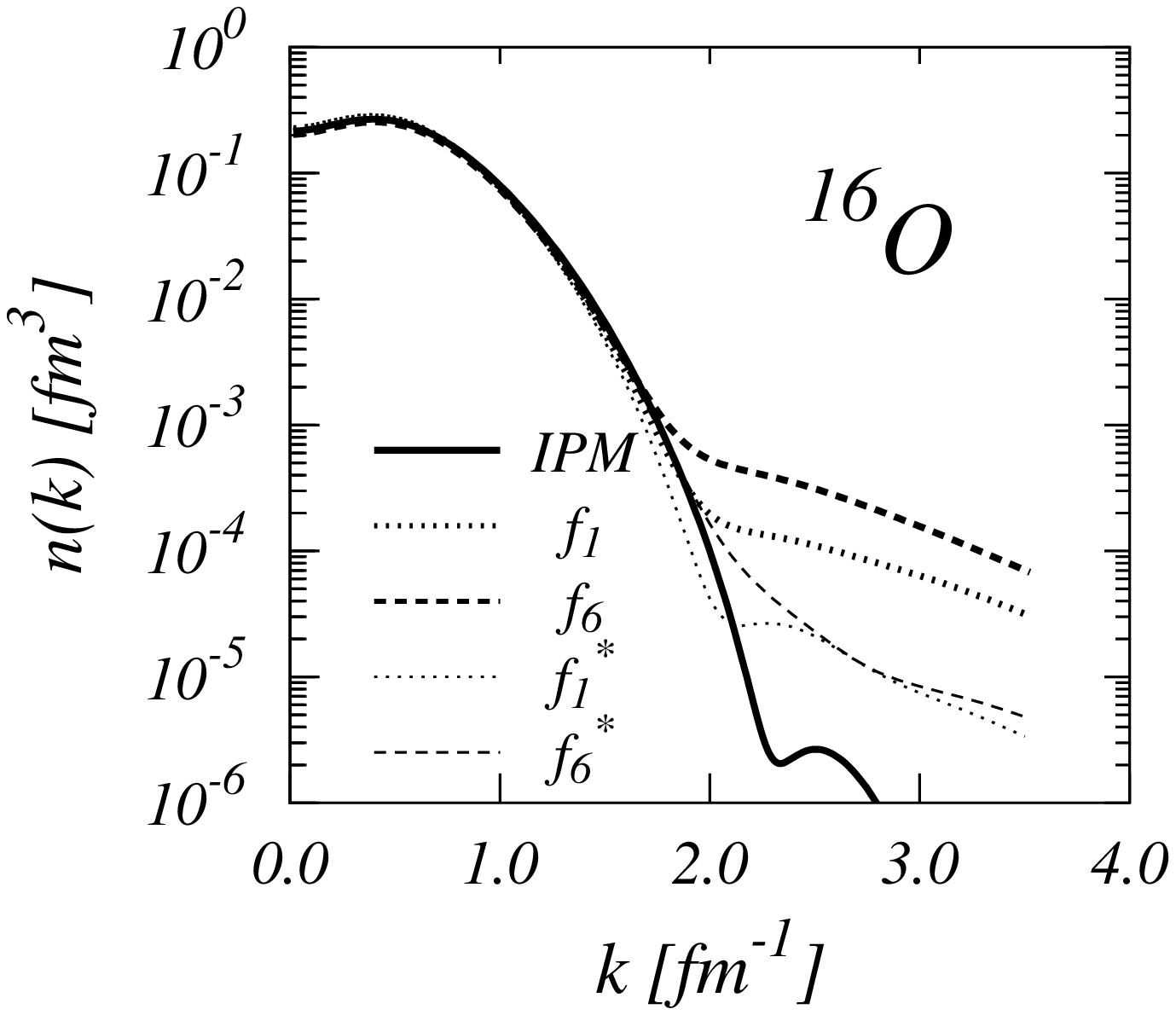}
\caption{\small 
Proton momentum distribution of \oxy in various approximations.
The thick lines are those of the analogous panel of Fig. 
\protect\ref{fig:md}. 
The thin lines have been obtained by using the first-order expansion
method of Ref. \protect\cite{ari97}.
}
\label{fig:mdcomp} 
\end{figure}

The general behavior of the momentum distributions, is very similar
for all the nuclei we have considered. Correlated and IPM
distributions almost coincide in the low momentum region up to a
precise value, when they start to deviate. The correlated
distributions show high momentum tails, which are orders of magnitude
larger than the IPM results.  The value of $k$ where uncorrelated and
correlated momentum distributions start to deviate, is smaller the
heavier is the nucleus.  It is about 1.9 fm$^{-1}$ for \car, and 1.5
fm$^{-1}$ for \lead. We notice that the value of the Fermi momentum of
symmetric nuclear matter at the saturation density is 1.36 fm$^{-1}$.

The results presented in Fig. \ref{fig:md} clearly show that the
effects of the scalar correlations are smaller than those obtained by
including the operator dependent terms.  We shall see in the following
that this is a common feature of our results.

Since relatively small differences are compressed in logarithmic
scale, we use the linear scale in Fig. \ref{fig:mdlin} to show, as
examples, the proton momentum distributions for \oxy and \lead nuclei,
multiplied by $k^2$.  This quantity, multiplied by a factor $4 \pi$,
is the probability of finding a proton with momentum $k$.  We observe
that the effects of the SRC on the quantity shown in Fig.
\ref{fig:mdlin} are basically two. The first one is the already
mentioned enhancement at large values of $k$. This effect is less
evident here than in Fig.  \ref{fig:md}.  The second effect, hardly
visible in Fig. \ref{fig:md}, is a reduction of the maxima which
appear approximately at $k$=1 fm$^{-1}$ in both nuclei.  These two
effects are obviously related, since all the momentum distributions
are normalized as indicated by Eq.  (\ref{eq:nor}), therefore
reductions and increases must compensate.

We found that the proton and neutron momentum distributions for nuclei
with $N=Z$ are very similar. For this reason we do not show the
neutron momentum distributions of the \car, \oxy and \caI nuclei.  We
compare in the panels (a) and (b) of Fig. \ref{fig:mdpn} the proton
and neutron momentum distributions of \caII and \lead.  The thicker
lines show the results of our FHNC/SOC calculation, the thinner lines
the IPM momentum distributions.

The figure shows that, in our calculations, the differences between
protons and neutrons momentum distributions are more related to the
different single particle structure than to the correlation effects.  
The main differences in the two distributions is in the zone where the
$n(k)$ values drops of orders of magnitudes. This corresponds to the
discontinuity region of the momentum distribution in the infinite
systems, which is related to the Fermi momentum. In a finite system, 
the larger number of neutrons implies that the neutron Fermi energy
is larger than that of the protons, and, consequently, the effective
Fermi momentum. For this reason, the neutrons momentum distributions
drops at larger values of $k$ than the proton distributions. 
In the panels (c) and (d) of Fig. \ref{fig:mdpn}, we show the 
quantity
\begin{equation}
\Delta = \frac{n^t_0(k)-n^t(k)}{n^t_0(k)+n^t(k)} \,\,,
\label{eq:delta}
\end{equation}
where $n^t_0(k)$ indicates the uncorrelated momentum distribution.
This quantity is useful to point out the effects of correlations. We
see that in the low $k$ region $\Delta$ is almost zero. After the
discontinuity region $\Delta$ reaches an almost constant value around
minus one.  This behavior indicates that in the low $k$ region the
momentum distribution is dominated by single particle dynamics.  The
differences between protons and neutrons at low $k$ are due to
different single particle wave functions.  In the higher $k$ region the
correlation plays an important role.  We observe that protons and
neutrons $\Delta$ are very similar, and this indicate that the effect
of the SRC is essentially the same for both kinds of nucleons. Our
results are in agreement with the findings of Ref.  \cite{boz04},
where $n(k)$ of asymmetric nuclear matter is presented.  There is
however a disagreement with the results of Ref. \cite{fri05}, where,
always in asymmetric nuclear matter, correlations effects between
protons were found to be stronger than those between neutrons.

The increase of the momentum distribution at large $k$ values, induced
by the SRC is a well known result in the literature, see for example
the review of Ref. \cite{ant88}.  The momentum distributions of
medium-heavy nuclei, have been usually obtained by using approximated
descriptions of the cluster expansion, which is instead considered at
all orders in our treatment.  The importance of a complete description
of the cluster expansion is exemplified in Fig.  \ref{fig:mdcomp},
where, together with our results, we also show the results of
Ref. \cite{ari97}, obtained by truncating the cluster expansion to the
first order in the correlation lines.  In both calculations the same
correlation functions and single particle basis, those of
Ref. \cite{bis06}, have been used.  The results obtained with the
first order approximation, provide only a qualitative description of
the correlation effects. They show high-momentum enhancements which,
however, underestimate the correct results by orders of magnitude.

\section{NATURAL ORBITS}
\label{sec:no}
The natural orbits are defined as those single particle wave functions
forming the basis where the OBDM is diagonal:
\begin{equation}
\rho^t(\br_1,\br'_{1})=\sum_{nlj}c_{nlj}^t\phi_{nlj}^{*\,t,NO}(\br_1)
\phi_{nlj}^{t,NO}(\br'_{1}) \,\,.
\label{eq:no}
\end{equation}
In the above equation the $c^t_{nlj}$ coefficients, called occupation
numbers, are real numbers.  In the IPM, the natural orbits correspond
to the mean-field wave functions of Eq.  (\ref{eq:spwf}), and the
$c^t_{nlj}$ numbers are 1, for the states below the Fermi surface, and
0 for those above it.

In order to obtain the natural orbits, we found convenient to express
the OBDM of Eq. (\ref{eq:no}) as:
\begin{equation}
\rho^t(\br_1,\br_{1'})=A^t(\br_1,\br_{1'})
\rho_o^t(\br_1,\br_{1'})+B^t(\br_1,\br_{1'}) \,\,,
\label{eq:rhosplit}
\end{equation}
where $\rho^t_{o}(\br_1,\br_{1'})$ is the uncorrelated OBDM of Eq. 
(\ref{eq:obdm0}), and the other two quantities are defined as:
\begin{eqnarray}
\nonumber
A^{t}(\br_1,\br'_1)&=&2C_{\omega,SOC}^{t}(\br_1)
C_{\omega,SOC}^{t}(\br'_1)
\exp[{N_{\omega\omega}^{t}(\br_1,\br'_1)}]
+\\ &&
2C_{\omega}^{t}(\br_1)C_{\omega}^{t}(\br'_1)
\exp[{N_{\omega\omega}^{t}(\br_1,\br'_1)}]\sum_{p>1}  
A^k\Delta^{k}N_{\omega\omega,p}^{t}(\br_1,\br'_1) \,\,,
\label{eq:arho}\\
\nonumber
B^{t}(\br_1,\br'_1)&=&-2C_{\omega,SOC}^{t}(\br_1)
C_{\omega,SOC}^{t}(\br'_1)
\exp[{N_{\omega\omega}^{t}(\br_1,\br'_1)}]
N_{\omega_c\omega_c}^{t}(\br_1,\br'_1)
- \\
\nonumber &&
2C_{\omega}^{t}(\br_1)C_{\omega}^{t}(\br'_1)
\exp{[N_{\omega\omega}^{t}(\br_1,\br'_1)]}\\
&&\times\sum_{p>1}  
A^k\Delta^{k} \Big\{ N_{\omega\omega,p}^{t}(\br_1,\br'_1) 
N_{\omega_c\omega_c}^{t}(\br_1,\br'_1)+
N_{\omega_c\omega_c,p}^{t}(\br_1,\br'_1)\Big\} \,\,.
\label{eq:brho}
\end{eqnarray}
The meaning of the $\omega\omega$, $\omega_c\omega_c$ labels used in
the above equations have been defined in \cite{fab01} where the index
$k$ has been defined as $p=2k+l-1$ with $l=0,1$ and $p=1,\ldots,6$.
The detailed expressions of the vertex corrections $C$ and of the
nodal functions $N$ are given in Appendix \ref{app:A}.

We expand the OBDM on a basis of spin spherical harmonics
${\bf Y}^m_{lj}$ , Eq. (\ref{eq:spwf}),
\begin{equation}
\rho^t(\br_1,\br'_{1}) = \sum_{ljm} \frac {1}{2j+1}
\left[{\cal A}_{lj}^t(r_1,r'_1) + {\cal B}_{lj}^t(r_1,r'_1) 
\right]
{\bf Y}^{* m}_{lj}(\Omega_1)  {\bf Y}^m_{lj}(\Omega'_1)
\label{eq:rhoexp}
\end{equation}
where $\Omega_1$ and $\Omega'_1$ indicate the polar angles identifying 
$\br_1$ and $\br'_1$.
The explicit expressions of the ${\cal A}$ and ${\cal B}$ coefficients
are:
\begin{eqnarray}
\nonumber
{\cal A}_{lj}^t(r_1,r'_1) &=&
(2l+1)\sum_{n_2l_2j_2l}(2l_2+1)(2j_2+1)
\left( \begin{array}{ccc}
       l & l_1 & l_2\\
       0 & 0 & 0 
      \end{array}\right)^2
\left\{ \begin{array}{ccc}
       j_2 & l_1 & j\\
       l & 1/2 & l_2 
      \end{array}\right\}^2 
\nonumber
\\
&~&R^t_{nl_2j_2}(r_1)R^t_{nl_2j_2}(r_2) A^t_{l_1}(r_1,r'_1)
\label{eq:cala}
\end{eqnarray}
with
\begin{equation}
A_l^t(r_1,r'_1)=\frac 2 {2l+1}
\int d\Omega A^t(\br_1,\br'_1)P_{l}(\cos\theta_{11'})
\end{equation}
and 
\begin{equation}
{\cal B}_{lj}^t(r_1,r'_1) 
=\frac{4\pi}{2l+1} \int 
d (\cos\theta_{11'}) B^t(\br_1,\br'_1)P_{l}(\cos\theta_{11'})
\label{eq:calb}
\end{equation}
In the above equations we have used the 3j and 6j Wigner symbols
\cite{edm57}.  The term ${\cal A}$ depends on both orbital and total
angular momenta of the single particle, $l$ and $j$ respectively, and
the term ${\cal B}$ depends only on the orbital angular momentum $l$.

As it has been done in Refs. \cite{lew88} and \cite{pol95} we 
identify the various natural orbit with a number, $\alpha$, ordering
them with respect to the decreasing value of the occupation
probability.  The general behavior of our results is analogous to
that described in Ref. \cite{lew88} where a system of $^3$He drops
composed by 70 atoms have been studied. The orbits corresponding to
states below the Fermi level in the IPM picture, have occupation
numbers very close to unity for $\alpha=1$, and very small in all the
other cases.
% occupation numbers protons
%
\begin{figure}[ht]
\includegraphics [scale=0.50]
                 {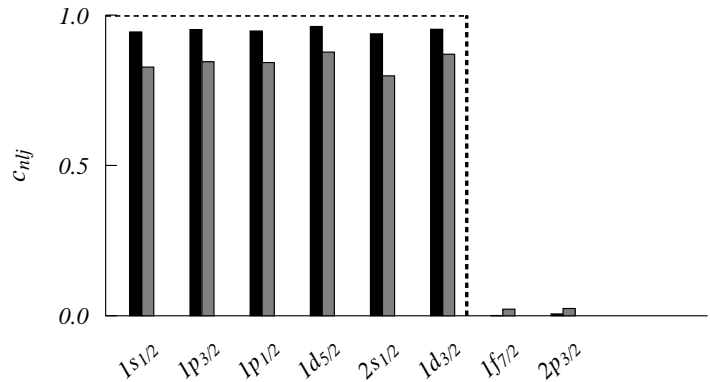}
\caption{\small Occupation numbers of the proton natural 
orbits of the \caII nucleus, having $\alpha=1$.
The dashed line indicates the IPM values. 
The black bars show the values obtained with the scalar correlation
and the gray bars those values obtained with the full correlation. 
}
\label{fig:natorb1} 
\end{figure}
%
% occupation numbers neutrons
%
\begin{figure}[ht]
\includegraphics [scale=0.50]
                 {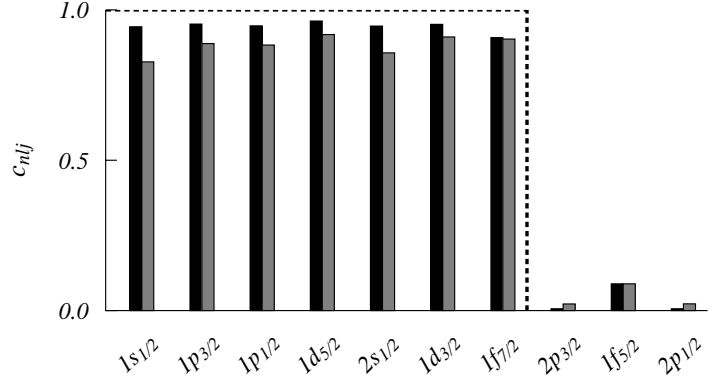}
\caption{\small The same as Fig. \protect\ref{fig:natorb1} for the 
occupation numbers of the neutron natural orbits of the
\caII nucleus. 
}
\label{fig:natorb2} 
\end{figure}
As example of our results, we show in Figs. \ref{fig:natorb1} and
\ref{fig:natorb2} the proton and neutron occupation numbers for the
natural orbits with$\alpha=1$ of the \caII nucleus.  In the
figures, the IPM results are indicated by the dashed lines. The black
bars show the values obtained by using scalar correlations only, the
gray bars those obtained with the complete operator dependent
correlations.

The correlated occupation numbers are smaller than one for orbits
below the Fermi surface, and larger than zero for those orbits above
the Fermi surface. This effect is enhanced by the operator dependent
correlations. We observe that, for the states above the Fermi surface,   
the gray bars are larger than the black ones, indicating that also
for these states the operator dependent correlations, produce larger
effects than the scalar ones.

\begin{figure}[pt]
\includegraphics [scale=0.50]
                 {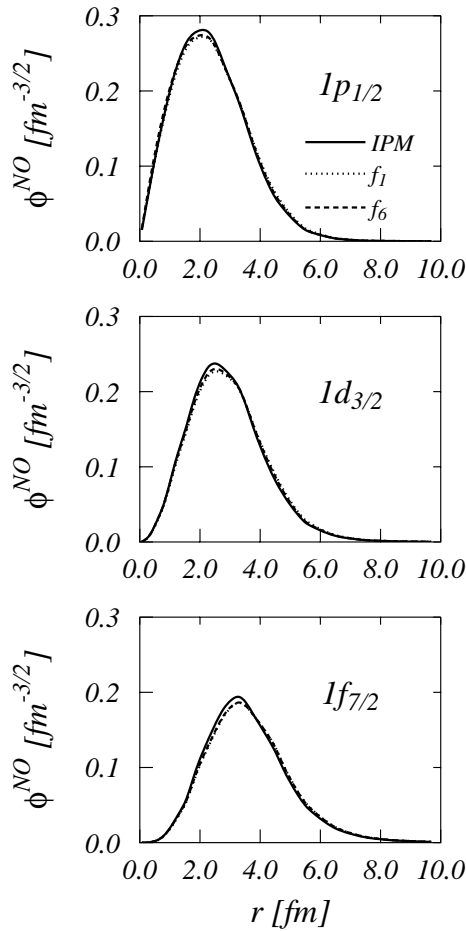}
\caption{\small Natural orbits for some neutron states in \caII.
 The full lines show the IPM orbits, the dotted lines those
 obtained with scalar correlations only and the dashed lines those
 obtained with the complete operator dependent correlation.
 }
\label{fig:natorb3} 
\end{figure}

We show in Fig. \ref{fig:natorb3} some $\alpha=1$ natural orbits for
three neutron states in \caII. In this figure, we compare the IPM
results (full lines) with those obtained with scalar correlation only
(dotted lines), and with the full operator dependent correlation
(dashed lines). The effect of the correlations is a lowering of the
peak and a small widening of the function. Despite the small effect,
it is interesting to point out that this is the only case where we
found that the inclusion of operator dependent terms diminishes the
effect of the scalar correlation.  This fact is coherent with the
results on the density distributions shown in Ref. \cite{bis06}.
\begin{table}[ht]
\begin{center}
\begin{ruledtabular}
\begin{tabular}{r|ccc|ccc|ccc}
\hline
 State       & & $\alpha=1$ & & &  $\alpha=2$ & & & $\alpha=3$ & \\  
\hline
             & $f_1$ & $f_6$ & PMD & $f_1$ & $f_6$ & PMD & 
             $f_1$ & $f_6$ & PMD \\ 
$1s_{1/2}$ (p)  & 0.956 & 0.873 & 0.921 & 0.011 & 0.038 & 0.013 
                & 0.002 & 0.007 & 0.002 \\
           (n)  & 0.957 & 0.873 &       & 0.012 & 0.039 & 
                & 0.003 & 0.008 &     \\
$1p_{3/2}$ (p)  & 0.973 & 0.921 & 0.947 & 0.004 & 0.013 & 0.007 
                & 0.001 & 0.003 & 0.001 \\
           (n)  & 0.973 & 0.924 &       & 0.004 & 0.014 & 
                & 0.002 & 0.004 &       \\
$1p_{1/2}$ (p)  & 0.970 & 0.923 & 0.930 & 0.003 & 0.012 & 0.008 
                & 0.001 & 0.003 & 0.002 \\
           (n)  & 0.970 & 0.922 &       & 0.004 & 0.013 & 
                & 0.002 & 0.003 &       \\
$1d_{5/2}$ (p)  & 0.001 & 0.005 & 0.016 & 0.013 & 0.003 & 0.003 
                & 0.000 & 0.000 & 0.000 \\
           (n)  & 0.001 & 0.005 &       & 0.001 & 0.003 & 
                & 0.000 & 0.000 &       \\
$1d_{3/2}$ (p)  & 0.002 & 0.005 & 0.019 & 0.001 & 0.003 & 0.005 
                & 0.000 & 0.000 & 0.001 \\
           (n)  & 0.001 & 0.005 &       & 0.001 & 0.003 & 
                & 0.000 & 0.000 &       \\
\hline
\end{tabular}
\end{ruledtabular}
\end{center}
\caption{\small Protons (p) and neutrons (n) natural orbits occupation 
numbers for \oxy. The PMD values are those of Ref. \cite{pol95}.
}
\label{tab:ocn}
\end{table}
In Tab. \ref{tab:ocn} we show the occupation numbers of the \oxy
protons and neutrons natural orbits also for $\alpha > 1$, and we make
a direct comparison with the results of Ref. \cite{pol95}. As already
said in the discussion of the \caII results, the inclusion of the
state dependent correlations increases the differences with respect to
the IPM. The occupation numbers of the orbits below the Fermi surface
are smaller than those obtained with scalar correlations only. The
situation is reversed for the orbits with $\alpha > 1$ or above the
Fermi level. For the states below the Fermi surface, our full
calculations produce correlation effects slightly larger than those
found in \cite{pol95}, whose results are closer to those we obtain
with scalar correlations only.  For orbits above the IPM Fermi
surface, our occupation numbers are always smaller than those of Ref.
\cite{pol95}.

\section{QUASI-HOLE WAVE FUNCTIONS AND THE SPECTROSCOPIC FACTORS}
\label{sec:specf}
The quasi-hole wave function is defined as:
\begin{equation}
\psi_{nljm}^{t}(x)=
\sqrt{A}\frac{<\Psi_{nljm}(A-1)|\delta(x-x_A)P^{t}_{A}|\Psi(A)>}
{\left[<\Psi_{nljm}(A-1)|\Psi_{nljm}(A-1)>
<\Psi(A)|\Psi(A)>\right]^{1/2}} \,\,,
\label{eq:qhfun}
\end{equation}
where $|\Psi_{nljm}(1,...,A-1)>$ and $|\Psi(1,...,A)>$ are the states
of the nuclei formed by $A-1$ and $A$ nucleons respectively, and
$P^t_A$ is the isospin projector.  In analogy to the ansatz
(\ref{eq:psi}), we assume that the state of the nucleus with $A-1$
nucleons can be described
as:
\begin{equation}
\Psi_{nljm}(A-1)=
{\cal F}(1,...,A-1)\Phi_{nljm}(1,...,A-1) \,\,,
\label{eq:psiam1} 
\end{equation}
where $\Phi_{nljm}(1,...,A-1)$ is the Slater determinant obtained by
removing from $\Phi(1,...,A)$ a particle characterized by the quantum
numbers $nljm$, and the correlation function ${\cal F}$
is formed, as indicated in Eq. (\ref{eq:cor2}), by the two-body
correlation functions $f_p$ obtained by minimizing the $A$ nucleon system.
In an uncorrelated system the quasi-hole wave
functions coincide with the hole mean-field wave functions
(\ref{eq:spwf}).

We are interested in the radial part of the quasi-hole wave function.
We obtain this quantity first 
by multiplying equation (\ref{eq:qhfun}) by the vector
spherical harmonics ${\bf Y}^{*m}_{lj}(\Omega)$, then, by integrating
over the angular coordinate $\Omega$, and, finally, 
by summing over $m$.  It is
useful to rewrite the radial part of the quasi-hole wave function as
\cite{fab01}:
\begin{equation}
\psi_{nlj}^{t}(r)= \frac 1 {2j+1}
\sum_m \int d\Omega \,
{\bf Y}^m_{lj}(\Omega)
\, \psi_{nljm}^{t}(x)
= \frac 1 {2j+1} \sum_m {\cal X}_{nljm}^t(r)[{\cal N}_{nlj}^t]^{1/2} \,\,,
\label{eq:qhprod}
\end{equation}
where we have defined:
\begin{equation}
{\cal{X}}_{nljm}^t(r)= 
\sqrt{A}\frac{<\Psi_{nljm}^t(A-1)|{\bf Y}^{*m}_{lj}(\Omega) \, 
\delta(\br-\br_{A})P^{t}_{A}|\Psi^t(A)>}{<\Psi_{nljm}^t(A-1)|
\Psi_{nljm}^t(A-1)>} \,\, ,
\end{equation}
and 
\begin{equation}
{\cal N}_{nljm}^t=<\frac{\Psi_{nljm}^t(A-1)|\Psi_{nljm}^t(A-1)>}
{<\Psi^t(A)|\Psi^t(A)>} \,\, .
\end{equation}
\begin{table}[ht]
\begin{center}
\begin{ruledtabular}
\begin{tabular}{c|ccc|ccc|ccc|ccc|ccc}
$nlj$ & &{\car} & & & {\oxy}& & &{\caI}& & &{\caII}& & &{\lead}\\ 
      &$f_1$ &$f_4$ &$f_6$&$f_1$ &$f_4$ &$f_6$&$f_1$ &$f_4$ &$f_6$&$f_1$ 
      &$f_4$ &$f_6$&$f_1$ &$f_4$ &$f_6$\\ 
$1s_{1/2}$ & 0.96  &0.95  &0.91 &0.95  &0.90  &0.85  &0.93  &0.84  &0.78  &  
    0.94  &0.85  &0.78  &0.93  &0.83  &0.77\\
$1p_{3/2}$  &0.96  &0.96  &0.94  &0.96  &0.93  &0.89  &0.95  &0.87  &0.82  &  
0.95  &0.87  &0.81  &0.94  &0.83 &0.77  \\
$1p_{1/2}$  &  &  &  &0.96  &0.93  &0.89  &0.95  &0.87  &0.81  &  
 0.95     &0.83  &0.80  &0.94  &0.83  &0.77 \\
$1d_{5/2}$  &  &  &  &  &  &  &0.96  &0.90  &0.86  &  
 0.96  &0.90  &0.85  &0.94  &0.84  &0.79\\
$2s_{1/2}$  &  &  &  &  &  &  &0.96  &0.92  &0.87  &  
0.94  &0.92  &0.86  &0.94  &0.86  &0.80\\
$1d_{3/2}$  &  &  &  &  &  &  &0.95  &0.90  &0.85  &  
 0.96  &0.90  &0.84  &0.94  &0.84  &0.79\\
$1f_{7/2}$  &  &  &  &  &  &  &  &  &  &  
      &  & & 0.94  &0.86  &0.81  \\
$2p_{3/2}$  &  &  &  &  &  &  &  &  &  &  
      &  & &0.95  & 0.87 &0.82\\
$1f_{5/2}$  &  &  &  &  &  &  &  &  &  &  
      &  &  &0.95  &0.86  &0.80\\
$2p_{1/2}$  &  &  &  &  &  &  &  &  &  &  
      &  &  &0.95  &0.87  &0.82\\
$1g_{9/2}$  &  &  &  &  &  &  &  &  &  &  
      &  &  & 0.95 &0.88  &0.83\\
$1g_{7/2}$  &  &  &  &  &  &  &  &  &  &  
      &  &  & 0.94 &0.88  &0.82\\
$2d_{5/2}$  &  &  &  &  &  &  &  &  &  &  
      &  &  &0.95  &0.89  &0.83\\
$1h_{11/2}$ &  &  &  &  &  &  &  &  &  &  
      &  &  & 0.94 &0.90  &0.86\\
$2d_{3/2}$  &  &  &  &  &  &  &  &  &  &  
      &  &  &0.95  &0.89  &0.83\\
$3s_{1/2}$  &  &  &  &  &  &  &  &  &  &  
      &  &  & 0.95 & 0.90 &0.85\\
\end{tabular}
\end{ruledtabular}
\end{center}
\caption{\small Proton spectroscopic factors of the \car, \oxy, 
\caI, \caII and \lead nuclei. 
We present the results obtained by using the scalar correlation only
($f_1$), the first four operator channels of the correlation 
($f_4$) and the full correlation operator ($f_6$).
}
\label{tab:sfp}
\end{table}
%% Spectroscopic factors
%

\begin{table}[ht]
\begin{center}
\begin{ruledtabular}
\begin{tabular}{c|ccc|ccc|ccc|ccc|ccc}
$nlj$ & &{\car} & & & {\oxy}& & &{\caI}& & &{\caII}& & &{\lead}\\ 
      &$f_1$ &$f_4$ &$f_6$&$f_1$ &$f_4$ &$f_6$&$f_1$ &$f_4$ &$f_6$&$f_1$ 
      &$f_4$ &$f_6$&$f_1$ &$f_4$ &$f_6$\\ 
$1s_{12}$ & 0.96  &0.95  &0.91 &0.95  &0.90  &0.85  &0.93  &0.84  &0.78  &  
  0.93  &0.86  &0.80  &0.92  &0.85  &0.80\\
$1p_{3/2}$  &0.96  &0.96  &0.94  &0.96  &0.93  &0.89  &0.95  &0.87  &0.82  &  
0.94  &0.88  &0.83  &0.93  &0.85 &0.80  \\
$1p_{1/2}$  &  &  &  &0.96  &0.93  &0.89  &0.95  &0.87  &0.81  &  
 0.94 &0.88  &0.82  &0.93  &0.85  &0.80 \\
$1d_{5/2}$  &  &  &  &  &  &  &0.96  &0.90  &0.86  &  
 0.95  &0.90  &0.86  &0.93  &0.86  &0.82\\
$2s_{1/2}$  &  &  &  &  &  &  &0.96  &0.92  &0.87  &  
0.95  &0.92  &0.87  &0.93  &0.88  &0.84\\
$1d_{3/2}$  &  &  &  &  &  &  &0.95  &0.90  &0.85  &  
 0.95  &0.90  &0.86  &0.93  &0.86  &0.82\\
$1f_{7/2}$  &  &  &  &  &  &  &  &  &  &  
  0.95    &0.94  &0.91 & 0.94  &0.88  &0.84  \\
$2p_{3/2}$  &  &  &  &  &  &  &  &  &  &  
      &  & &0.94  & 0.89 &0.85\\
$1f_{5/2}$  &  &  &  &  &  &  &  &  &  &  
      &  &  &0.93  &0.88  &0.84\\
$2p_{1/2}$  &  &  &  &  &  &  &  &  &  &  
      &  &  &0.94  &0.89  &0.85\\
$1g_{9/2}$  &  &  &  &  &  &  &  &  &  &  
      &  &  & 0.94 &0.90  &0.86\\
$1g_{7/2}$  &  &  &  &  &  &  &  &  &  &  
      &  &  & 0.94 &0.90  &0.86\\
$2d_{5/2}$  &  &  &  &  &  &  &  &  &  &  
      &  &  &0.94  &0.91  &0.87\\
$1h_{11/2}$ &  &  &  &  &  &  &  &  &  &  
      &  &  & 0.94 &0.93  &0.89\\
$2d_{3/2}$  &  &  &  &  &  &  &  &  &  &  
      &  &  &0.94  &0.90  &0.87\\
$3s_{1/2}$  &  &  &  &  &  &  &  &  &  &  
      &  &  & 0.94 & 0.92 &0.88\\
$2f_{7/2}$  &  &  &  &  &  &  &  &  &  &  
      &  &  &0.95  &0.93  &0.90\\
$1h_{9/2}$ &  &  &  &  &  &  &  &  &  &  
      &  &  & 0.94 &0.92  &0.88\\
$2f_{5/2}$  &  &  &  &  &  &  &  &  &  &  
      &  &  &0.95  &0.93  &0.90\\
$3p_{3/2}$  &  &  &  &  &  &  &  &  &  &  
      &  &  & 0.95 & 0.94 &0.90\\
$1i_{13/2}$  &  &  &  &  &  &  &  &  &  &  
      &  &  &0.94  &0.93  &0.90\\
$3p_{1/2}$ &  &  &  &  &  &  &  &  &  &  
      &  &  & 0.95 &0.94  &0.90\\
\end{tabular}
\end{ruledtabular}
\end{center}
\caption{\small The same as Tab. \protect\ref{tab:sfp}
for neutron states.
}
\label{tab:sfn}
\end{table}

Following the procedure outlined in Ref.  \cite{fab01}, we consider
separately the cluster expansions of the two terms ${\cal N}_\alpha^t$ and
${\cal X}_\alpha^t$, where we have indicated with $\alpha$ the set of 
the $nljm$ quantum numbers. 
We obtain for ${\cal X}_\alpha^t$ the expression:
\begin{eqnarray}
\nonumber
{\cal X}_{\alpha}^{t}(r)&=&
C^{t,\alpha}_{\omega,SOC}(\br)
\Bigg(
R^{t}_{nlj}(r)+
\int d^{3}r_1R^{t}_{nlj}(r_1)P_{l}(\cos{\theta}) \\
\nonumber &~& \times
\Bigg\{
g_{\omega d}^{tt,\alpha}(\br,\br_1)C_{d,pq}^{t,\alpha}(\br_1)
\Big[
-\rho_{o}^{t,\alpha}(\br,\br_1)+N_{\omega_cc}^{t,\alpha}(\br,\br_1)
\Big]
 \\ 
%\nonumber 
&~&
+\rho_{o}^{t,\alpha}(\br,\br_1)
-N^{t,\alpha}_{\omega_c\rho}(\br,\br_1)-N^{t,\alpha}_{\rho\rho}(\br,\br_1)
%\\ &&
+{\cal X}_{SOC}^{t}(\br,\br_1)
\Bigg\}
\Bigg) \,\,,
\end{eqnarray}
and for ${\cal N}_\alpha^t$ the expression: 
\begin{eqnarray}
\nonumber
\Big[{\cal N}_{\alpha}^{t}\Big]^{-1}&=&\int
d^3rC_{d,pq}^{t,\alpha}(\br)
\Bigg(
|\phi^{t}_{\alpha}(\br)|^2+
\int d^3r_1\phi^{t *}_{\alpha}(\br)\phi^{t}_{\alpha}(\br_1)
\\ \nonumber &~& \times
2
\Big\{
g_{dd}^{tt,\alpha}(\br,\br_1)C_{d,pq}^{t,\alpha}(\br_1)
\Big[-\rho_{o}^{t,\alpha}(\br,\br_1)+N^{t,\alpha}_{cc}
(\br,\br_1)\Big]\\
%\nonumber
&& +\rho_{o}^{t,\alpha}(\br,\br_1)-N_{x\rho}^{t,\alpha}(\br,\br_1)
-N_{\rho\rho}^{t,\alpha}(\br,\br_1)
%\\
%&&
+{\cal N}_{SOC}^{t}(\br,\br_1)\Big\}
\Bigg) \,\,,
\end{eqnarray} 

The expressions of the functions 
${\cal N}_{SOC}^{t}(\br,\br_1)$, 
${\cal X}_{SOC}^{t}(\br,\br_1)$, are:
\begin{eqnarray}
\nonumber
{\cal X}_{SOC}^{t_{1}}(\br,\br_1)&=&
\sum_{k=1}^{3}A^k\sum_{t_{2}=p,n}
\Big[(1-\delta_{k,1}){\cal X}_{2k-1,2k-1}^{t_{1}t_{2}}(\br,\br_1)\\
&&+\chi_{1}^{t_{1}t_{2}}\Big({\cal X}_{2k-1,2k}^{t_{1}t_{2}}(\br,\br_1)
+{\cal X}_{2k,2k-1}^{t_{1}t_{2}}(\br,\br_1)\Big)+
\chi_{2}^{t_{1}t_{2}}{\cal X}_{2k,2k}^{t_{1}t_{2}}(\br,\br_1)\Big]
\,\,,\\
\nonumber
{\cal N}_{SOC}^{t_{1}}(\br,\br_1)&=&
\sum_{k=1}^{3}A^k\sum_{t_{2}=p,n}
\Big[(1-\delta_{k,1}){\cal N}_{2k-1,2k-1}^{t_{1}t_{2}}(\br,\br_1)\\
&&+\chi_{1}^{t_{1}t_{2}}\Big({\cal N}_{2k-1,2k}^{t_{1}t_{2}}(\br,\br_1)
+{\cal X}_{2k,2k-1}^{t_{1}t_{2}}(\br,\br_1)\Big)+
\chi_{2}^{t_{1}t_{2}}{\cal
N}_{2k,2k}^{t_{1}t_{2}}(\br,\br_1)\Big]
\,\,.
\end{eqnarray}
where the indexes $t$ refer to the isospin, and we have defined:
\begin{eqnarray}
\nonumber
{\cal X}_{pq}^{t_{1}t_{2}}(\br,\br_1)&=&\frac{1}{2}
\Big[h_{\omega,p}^{t_{1}t_{2} ,\alpha}(\br,\br_1)
g_{\omega d}^{t_{1}t_{2} ,\alpha}(\br,\br_1)
C_{d}^{t_{2},\alpha}(\br_1)\Big(-\rho_{o}^{t_{2},\alpha}(\br,\br_1)+
N_{\omega_cc}^{t_{2},\alpha}(\br,\br_1)\Big)\\
&&+g_{\omega d}^{t_{1}t_{2},\alpha}(\br,\br_1)C^{t_{2},\alpha}_{d}(\br_1)
N_{\omega_cc,p}^{t_{2},\alpha}(\br,\br_1)-
N_{\omega_c\rho,p}^{t_{2},\alpha}(\br,\br_1)-
N_{\rho\rho,p}^{t_{2},\alpha}(\br,\br_1)\Big]\Delta^{k_2}
\,\,, \\
\nonumber
{\cal N}_{pq}^{t_{1}t_{2}}(\br,\br_1)&=&
\Big[h_{d,p}^{t_{1}t_{2} ,\alpha}(\br,\br_1)g_{dd}^{t_{1}t_{2} ,\alpha}
(\br,\br_1)
C_{d}^{t_{2},\alpha}(\br_1)\Big(-\rho_{o}^{t_{2},\alpha}(\br,\br_1)+
N_{cc}^{t_{2},\alpha}(\br,\br_1)\Big)\\
&&+g_{dd}^{t_{1}t_{2}, \alpha}(\br,\br_1)C_{d}^{t_{2},\alpha}(\br_1)
N^{t_{2},\alpha}_{cc,p}(\br,\br_1)-
N_{x\rho,p}^{t_{2},\alpha}(\br,\br_1)-
N_{\rho\rho,p}^{t_{2},\alpha}(\br,\br_1)\Big]\Delta^{k_2}
\,\, ,
\end{eqnarray}
where $k_2=1,2,3$ for $q=1,3,5$.
The expressions of the other terms 
are given in Appendix \ref{app:A}. 
All the quantities used in the above expressions depend on the set
of quantum numbers $\alpha$ characterizing the quasi-hole state, 
since we have written the various equations by using 
\cite{ari01}:
\begin{equation}
\rho_{o}^{t,\alpha}(\br,\br_1)=\rho_{o}^{t}(\br,\br_1)-
\phi^{t *}_{\alpha}(\br)\phi^{t}_{\alpha}(\br_1) \, \,.
\end{equation}
The knowledge of the quasi-hole functions allows us to calculate the
spectroscopic factors:
\begin{equation}
S_{nlj}^t=\int dr\,r_1^2\,|\psi_{nlj}^t(r)|^2
\,\,.
\label{eq:sf}
\end{equation}
%
%
% natural orbits 
%
%
% quasi-hole wave functions
%
\begin{figure}[pt]
\includegraphics[scale=0.50]
                {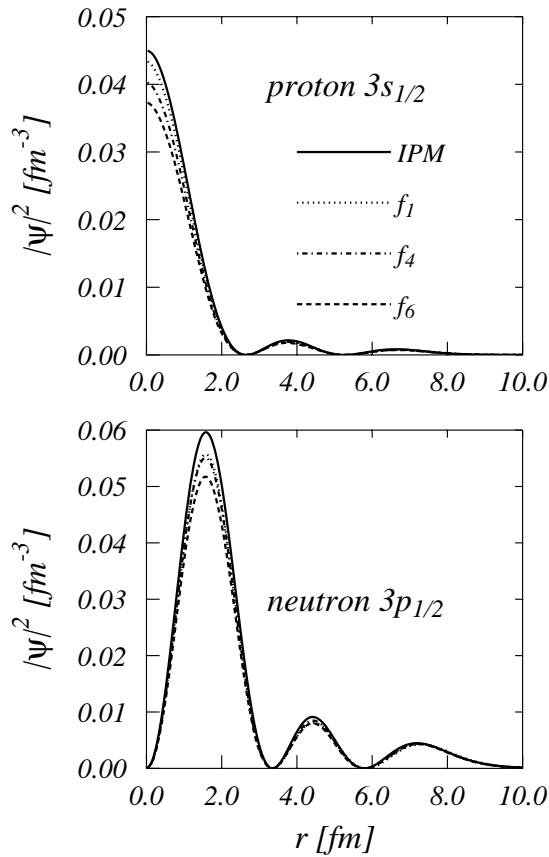} 
\caption{\small 
  Proton $3s_{1/2}$ and neutron $3p_{1/2}$ quasi-hole functions,
  squared, of the \lead nucleus.  The various lines show the results
  obtained by using different type of correlations.  }
\label{fig:qh_fun}
\end{figure}
%
% experimental thallium difference
%
\begin{figure}[pt]
\includegraphics[scale=0.50] 
               {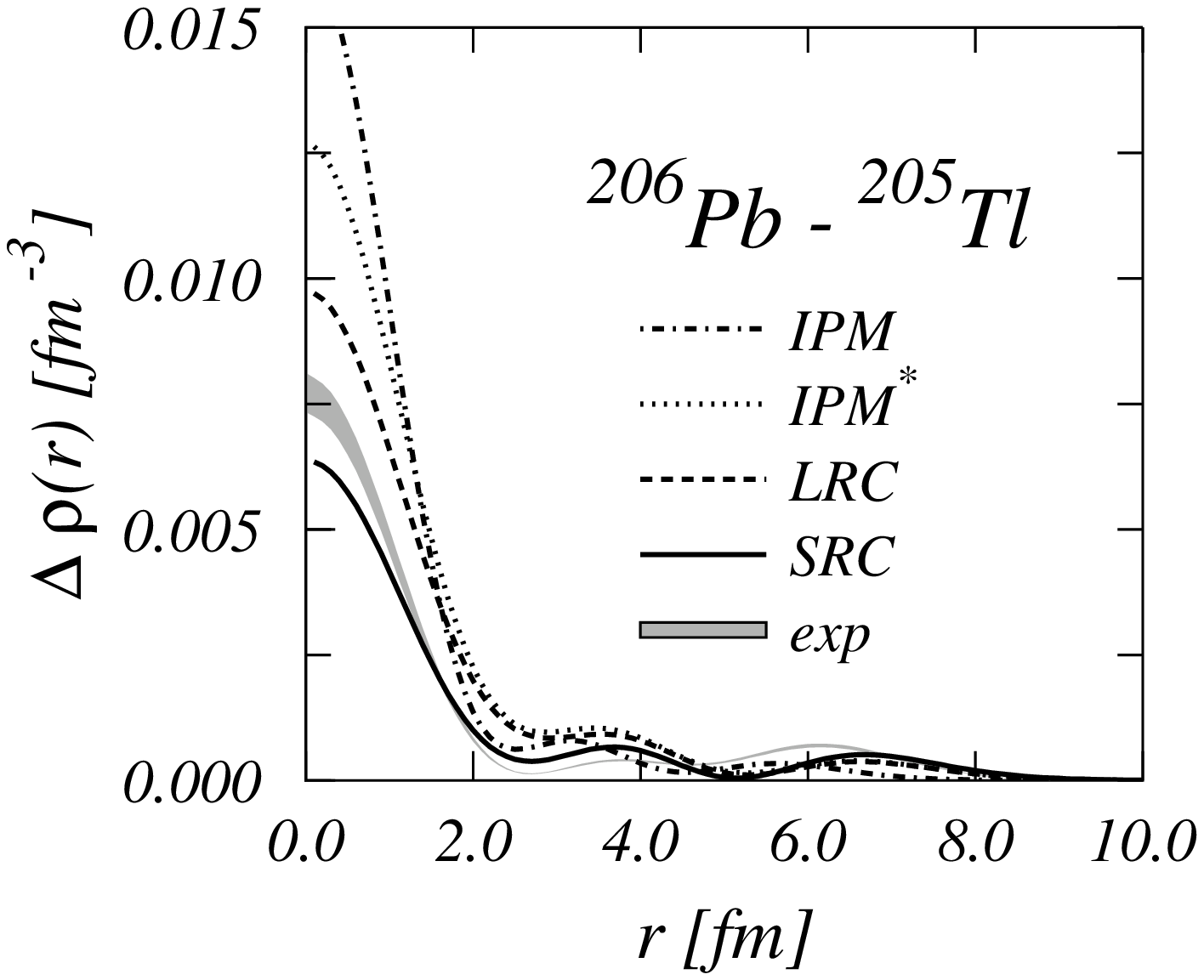} 
\caption{\small Differences between charge density distributions of 
$^{206}$Pb and $^{205}$Tl. See the text for the explanation of the
various lines. 
}
\label{fig:205tl}
\end{figure}
The proton and neutron spectroscopic factors for the \car, \oxy, \caI,
\caII and \lead nuclei are given in Tabs. \ref{tab:sfp} and
\ref{tab:sfn} for each single hole state.  In these tables we compare
the results obtained by using scalar correlations ($f_1$), with those
obtained with the four central channels ($f_4$) and with the full
correlation ($f_6$).  The inclusion of the correlations produce
spectroscopic factors smaller than one, the mean-field value.  The
$f_6$ results are smaller than those of $f_4$, which are smaller than
those obtained with $f_1$.

We found that the effect of the correlations becomes larger the more
bound is the state.  This fact emerges by observing that for a fixed
set of $lj$ quantum numbers the spectroscopic factors increase with
$n$, and, at the same time, that the values of the spectroscopic
factors become larger when $n$ and the $lj$ values increase.

The values of the spectroscopic factors depend on the choice of the
single particle basis. As we have already said in the introduction,
our results have been obtained by using the Woods-Saxon single
particle bases given in Ref. \cite{bis06}. This basis has been chosen
in order to reproduce the single particle energies around the Fermi
surface and the charge distribution of each nucleus considered. The
correlation function has been fixed by the minimization procedure
(\ref{eq:varprin}). We tested the sensitivity of our results to
different single particle basis, by calculating \oxy and \caI
spectroscopic factors by using with the Harmonic Oscillator and
Woods-Saxon single particle wave functions of Ref. \cite{fab00}, fixed
by a global minimization of the energy. Despite the remarkable
differences between the various single particle basis, we found that
the maximum variations in the values of the spectroscopic factors is
of about the 5\%.  This value is smaller than the variations produced
by the different terms of the correlations, shown in Tabs.
\ref{tab:sfp} and \ref{tab:sfn}.  This indicates that our results are
more sensitive to the SRC than to the choice of the single particle
basis.
 
As example of correlation effects on the quasi-hole wave functions, we
 show in Fig. \ref{fig:qh_fun} the squares of the proton $3s_{1/2}$
 and neutron $3p_{1/2}$ quasi-hole wave functions for the \lead
 nucleus.  The correlations lower the wave function in the nuclear
 interior. Also in this case, the effect of the correlations increases
 together with the number of operator channels considered.

In Fig. \ref{fig:205tl} we show with a gray band the difference
between the empirical charge distributions of $^{206}$Pb and
$^{205}$Tl \cite{cav82}. The dashed dotted line, labeled as IPM, has
been obtained by considering that the difference between the two
charge distributions can be described as a single $3s_{1/2}$ proton
hole in the core of the lead nucleus.  This curve has been obtained by
folding the IPM result of Fig. \ref{fig:qh_fun} with the electric proton
form factor in its dipole form.  In a slightly more elaborated
picture, the ground state of the $^{205}$Tl is composed by the
$3s_{1/2}$ proton hole in the $^{206}$Pb ground state, plus the
coupling of the $2d_{5/2}$ and $2d_{3/2}$ proton levels with the first
$2^+$ excited state of $^{206}$Pb \cite{zam75,kle76}.  This
description of the $^{205}$Tl charge distribution, shown by the dotted
line in the figure, is still within the IPM framework. The dashed line
has been obtained by adding to the dotted line the core polarization
effects produced by long-range correlations. These effects have been
calculated by following the Random Phase Approximation approach of
Refs. \cite{co87c,ang01}. The full line has been obtained when our SRC
effects are also included.

The various effects presented in Fig. \ref{fig:205tl} have been
obtained in different theoretical frameworks, and the final result
does not have any pretense of being a well grounded and coherent
description of the empirical charge differences.  The point we want to
make by showing this figure is that the effects of the SRC are of the
same order of magnitude of those commonly considered in traditional
nuclear structure calculations.

\section{SUMMARY AND CONCLUSIONS}
\label{sec:summary}

In this work we have extended the FHNC/SOC scheme in order to
calculate the OBDM's, the natural orbits and the quasi-hole wave
functions of finite nuclear systems non saturated in isospin, and in
$jj$ coupling representation of the single particle wave function
basis.  Our results have been obtained by using the many-body wave
functions obtained by minimizing the nuclear hamiltonian with the
two-body realistic interaction Argonne $v'_8$ and the associated
three-body interaction Urbana IX. The calculations have been done by
using operator dependent correlations which include terms up to the
tensor ones. 

We found that the correlations enhance by orders of magnitude the
high-energy tail of the nucleon momentum distribution. The occupation
numbers of the natural orbits below the Fermi level, are depleted, and
the opposite happens for those above the Fermi level. Also the values
of the spectroscopic factors are depleted with respect to the IPM.  A
reliable comparison between our spectroscopic factors with the
empirical ones requires the description of the reactions used to
extract them, and this is part of our future projects.

We have shown that the results of models considering expansions up to
the first order correlation lines, provide only qualitative
descriptions of the SRC effects.  In the description of the charge
density difference between $^{206}$Pb and $^{205}$Tl, the SRC effects
are of comparable size of those commonly considered in traditional
nuclear structure calculations based on effective theories.

A general outcome of our study, is that the effects of the
correlations increase with the complexity of the correlation function.
This means that operator dependent correlations enhance the effects
produced by the scalar correlations. This not obvious result, is valid
in general, not always. We have shown in Ref. \cite{bis06}, that
scalar and operator dependent correlations have destructive
interference effects on the density distributions.  We found in the
present study an analogous behavior regarding the natural orbits.
These quantities are related to the density distributions.  It seems
that the effects of the SRC are rather straightforward on quantities
which involve two-nucleons, while they are more difficult to predict
on quantities related to single nucleon dynamics. On these last
quantities, however, these SRC effects are very small, usually
negligible.

In our calculations the nuclear interaction acts only in defining the
many-body wave functions by means of the variational principle
(\ref{eq:varprin}), more specifically, in selecting the correlation
function (\ref{eq:cor2}). It is therefore difficult to disentangle the
role played by the various parts of the interaction, e.g. the tensor
part of the three-body force, on the quantities we have studied in
this article. We have instead evaluated the effects of the various
parts of the correlation function.

In this work, we have highlighted a set of effects that cannot be
described by mean field based effective theories. The description of
the nucleus in kinematics regimes where these effects are relevant,
requires the use of microscopic theories.

\section{ACKNOWLEDGMENTS}
This work has been partially supported by the agreement INFN-CICYT, by
the Spanish Ministerio de Educaci\'on y Ciencia (FIS2005-02145) 
and by the MURST through the PRIN: {\sl Teoria della struttura dei nuclei 
e della materia nucleare}.

%-----------------------------------------
%  appendices
%-----------------------------------------
\appendix
\section{}
\label{app:A}

For sake of completeness, we give in this appendix the detailed
expression of the OBDM for finite nuclear systems not saturated in
isospin, and in $jj$ coupling scheme of the single particle wave
function basis (\ref{eq:spwf}).  The notation for the nodal functions
$N$ and for the vertex corrections $C$ is that used in Ref.
\cite{fab01}. The indexes $t_{1},t_{2},t_{3}$ indicate protons and
neutrons, and the subscript $j$ is related to the antiparallel spin
states.

For the correlated OBDM we obtain the expression:
\begin{eqnarray}
\rho^{t}(\br_1,\br_{1'})&=&2C_{\omega,SOC}^{t}(\br_1)
C_{\omega,SOC}^{t}(\br_{1'})
e^{N_{\omega\omega}^{t}(\br_1,\br_{1'})}
\Big[\rho_{o}^{t}(\br_1,\br_{1'})-
N_{\omega_c\omega_c}^{t}(\br_1,\br_{1'})\Big]
+ \\
\nonumber &&
2C_{\omega}^{t}(\br_1)C_{\omega}^{t}(\br_{1'})
e^{N_{\omega\omega}^{t}(\br_1,\br_{1'})}\\
\nonumber
&&\times\sum_{p>1}  
A^k\Delta^{k} \Big\{ N_{\omega\omega,p}^{t}(\br_1,\br_{1'})
\Big[\rho_{o}^{t}(\br_1,\br_{1'})-
N_{\omega_c\omega_c}^{t}(\br_1,\br_{1'})\Big]-
N_{\omega_c\omega_c,p}^{t}(\br_1,\br_{1'})\Big\} \,\,.
\label{eq:obdm2}
\end{eqnarray}
In the above equation, $k$ has been defined as in Eq. (\ref{eq:brho}), 
and we have used
$\Delta^{k=1,2,3}=1-\delta_{k,3}$, and 
$A^{k=1,2,3}=1,3,6$. 

In the following we shall calculate the 
expectation value of the isospin operator sequence:
\[
\chi_n^{\ton \ttw} = \chi_{\ton}^* (1)\chi_{\ttw}^* (2)
\left(\btau_1 \cdot \btau_2 \right)^n \chi_{\ton} (1)
\chi_{\ttw} (2) \,\,,
\]
by considering that 
\[
\left(\btau_i \cdot \btau_j \right)^n = a_n + 
(1-a_n) \btau_i \cdot \btau_j \,\,,
\]
with 
\[
a_{n+1}=3(1-a_n) \hspace {1.0 cm} {\rm and} \hspace {1.0 cm} a_0=1  
\,\,.
\]
By using the above equations we have that:
\[
\chi_0^{\ton \ttw}= 1 
\hspace {0.5 cm} {\rm ,} \hspace {0.5 cm}
\chi_1^{\ton \ttw}= 2\delta_{\ton \ttw}-1
\hspace {0.5 cm} {\rm and} \hspace {0.5 cm}
\chi_n^{\ton \ttw} = 2a_n-1 +2(1-a_n) \delta_{\ton \ttw}
\,\,.
\]

The expressions of the vertex corrections are:
\begin{equation}
C_{\omega,SOC}^{t_{1}}(\br_1)=C_{\omega}^{t_{1}}(\br_1)\Big[1+
U_{\omega,SOC}^{t_{1}}(\br_1)\Big]
\,\,,
\end{equation}
\begin{eqnarray}
\nonumber
U^{t_{1}}_{\omega,SOC}(\br_1)&=&\sum_{k=1}^{3}A^k\sum_{t_{2}=p,n}
\Big[(1-\delta_{k,1})U_{\omega,2k-1,2k-1}^{t_{1}t_{2}}(\br_1)\\
&&+\chi_{1}^{t_{1}t_{2}}\Big(U_{\omega,2k-1,2k}^{t_{1}t_{2}}(\br_1)
+U_{\omega,2k,2k-1}^{t_{1}t_{2}}(\br_1)\Big)+
\chi_{2}^{t_{1}t_{2}}U_{\omega,2k,2k}^{t_{1}t_{2}}(\br_1)\Big]
\,\,,
\end{eqnarray}
where we have defined 
\begin{eqnarray}
\nonumber
U^{t_{1}t_{2}}_{\omega,pq}(\br_1)&=&\int
d\br_2 h_{\omega,p}^{t_1 t_2}(r_{12})\Bigg\{\Big[g_{\omega
d}^{t_{1}t_{2}}(\br_1,\br_2)C_{d,pq}^{t_{2}}(\br_2)+g_{\omega
e}^{t_{1}t_{2}}(\br_1,\br_2)
C_{e,pq}^{t_{2}}(\br_2)\Big]N^{t_{1}t_{2}}_{\omega d,q}(\br_1,\br_2)\\
&& +g^{t_{1}t_{2}}_{\omega d}(\br_1,\br_2)C_{e,pq}^{t_{2}}(\br_2)
N_{\omega e,q}^{t_{1}t_{2}}(\br_1,\br_2)\Bigg\}
\,\,,
\\
h^{t_{1}t_{2}}_{\omega,p}(\br_1,\br_2)& = &
\frac{f_{p}(r_{12})}{f_1(r_{12})}+
(1-\delta_{p,1})N_{\omega d,p}^{t_{1}t_{2}}(\br_1,\br_2)
\,\,.
\end{eqnarray}

The expressions of the 
two-body distribution functions for $p>1$ are:
\begin{eqnarray}
g_{\omega\omega,p}^{t_{1}}(\br_1,\br_{1'})&=&
g^{t_{1}}_{\omega\omega}(\br_1,\br_{1'})
N^{t_{1}}_{\omega\omega,p}(\br_1,\br_{1'}) \,\,,
\\
g_{\omega d,p}^{t_{1}t_{2}}(\br_1,\br_2)&=&
\nonumber
h^{t_{1}t_{2}}_{\omega,p}(\br_1,\br_2)g_{\omega
d}^{t_{1}t_{2}}(\br_1,\br_2) \\
&=&N^{t_{1}t_{2}}_{\omega d,p}(\br_1,\br_2)+X^{t_{1}t_{2}}_{\omega
d,p}(\br_1,\br_2) \,\,,
\\
\nonumber
g^{t_{1}t_{2}}_{\omega e,p}(\br_1,\br_2)&=&
h^{t_{1}t_{2}}_{\omega,p}(\br_1,\br_2)
g_{\omega e}^{t_{1}t_{2}}(\br_1,\br_2)+g_{\omega
d}^{t_{1}t_{2}}(\br_1,\br_2)N^{t_{1}t_{2}}_{\omega
e,p}(\br_1,\br_2) \\
&=&X^{t_{1}t_{2}}_{\omega e,p}(\br_1,\br_2)+N^{t_{1}t_{2}}_{\omega
e,p}(\br_1,\br_2)
\,\,,
\\
\nonumber
g_{\omega_cc,p}^{t_{1}}(\br_1,\br_2)&=&
h^{t_{1}t_{1}}_{\omega,p}(\br_1,\br_2)
g_{\omega_cc}^{t_{1}}(\br_1,\br_2)
+g_{\omega d}^{t_{1}t_{1}}(\br_1,\br_2)N_{\omega_cc,p}^{t_{1}}
(\br_1,\br_2) \\
&=&X^{t_{1}}_{\omega_c c,p}(\br_1,\br_2)+N^{t_{1}}_{\omega_c c,p}
(\br_1,\br_2)
\,\,.
\end{eqnarray}

Finally the nodals functions are expressed as:
\begin{eqnarray}
N_{mn(j),p}^{t_{1}}(1,2)&=&N_{mnx(j),p}^{t_{1}}(1,2)+
N_{m\rho(j),p}^{t_{1}}(1,2)
+N_{\rho n(j),p}^{t_{1}}(1,2)+N_{\rho,p}^{t_{1}}(1,2)
\,\,,
\end{eqnarray}
with $m,n=c,w_c$.  The separation of the above nodal diagrams in four
terms, corresponds to the classification in the $xx$, $x \rho$, $\rho
x$ and $\rho \rho$ parts \cite{bis06}, and it has been applied to the
quantities $N_{mn(j),pqr}^{t_{1}t_{3}}(1,2)$ defined in the following.
\begin{eqnarray}
\nonumber 
N_{mm,2k_{1}-1}^{t_{1}}(1,1')&=&
     \sum_{k_{2}k_{3}=1}^{3}\sum_{t_{3}=p,n}
     \Big[N_{mm,2k_{1}-1,2k_{2}-1,2k_{3}-1}^{t_{1}t_{3}}(1,1')\\
&+& \chi_{1}^{t_{1}t_{3}}\big[
          N_{mm,2k_{1}-1,2k_{2},2k_{3}-1}^{t_{1}t_{3}}(1,1')+ 
N_{mm,2k_{1}-1,2k_{2}-1,2k_{3}}^{t_{1}t_{3}}(1,1')\big]\Big]
\,\,,
\\
N_{mm,2k_{1}}^{t_{1}t_{2}}(1,1')&=&\sum_{k_{2},k_{3}=1}^{3}
          \sum_{t_{3}=p,n}\chi^{t_{1}t_{3}}_2
           N_{mm,2k_{1},2k_{2},2k_{3}}^{t_{1}t_{3}}(1,1')
\,\,,
\\
N_{\omega n,2k_{1}-1}^{t_{1}t_{2}}(1,2)&=&
     \sum_{k_{2}k_{3}=1}^{3}\sum_{t_{3}=p,n}
  \Big[N_{\omega n,2k_{1}-1,2k_{2}-1,2k_{3}-1}^{t_{1}t_{2}t_{3}}(1,2)\\
\nonumber &+ &\chi_{1}^{t_{1}t_{3}}
          N_{\omega n,2k_{1}-1,2k_{2},2k_{3}-1}^{t_{1}t_{2}t_{3}}(1,2)
          +\chi_{1}^{t_{2}t_{3}}
  N_{\omega n,2k_{1}-1,2k_{2}-1,2k_{3}}^{t_{1}t_{2}t_{3}}(1,2)\Big]
\,\,,
\\
N_{\omega n,2k_{1}}^{t_{1}t_{2}}(1,2)&=&\sum_{k_{2},k_{3}=1}^{3}
          \sum_{t_{3}=p,n}N_{\omega
n,2k_{1},2k_{2},2k_{3}}^{t_{1}t_{2}t_{3}}(1,2)
\,\,,
\\
N_{\omega_cc(j),2k_{1}-1}^{t_1}(1,2)&=&
 \sum_{k_{2},k_{3}=1}^{3}\sum_{t_{3}=p,n}
\Big[N_{\omega_cc(j),2k_{1}-1,2k_{2}-1,2k_{3}-1}^{t_{1}t_{3}}(1,2)\\
\nonumber &+&\chi_{1}^{t_{1}t_{3}}
  \big[N_{\omega_cc(j),2k_{1}-1,2k_{2},2k_{3}-1}^{t_{1}t_{3}}(1,2)+
N_{\omega_cc(j),2k_{1}-1,2k_{2}-1,2k_{3}}^{t_{1}t_{3}}(1,2)\big]\Big]\\
N_{\omega_cc(j),2k_{1}}^{t_{1}}(1,2)&=&
 \sum_{k_{2},k_{3}=1}^{3}\sum_{t_{3}=p,n}
  N_{\omega_cc(j),2k_{1},2k_{2},2k_{3}}^{t_{1}t_{3}}(1,2)
\,\,.
\end{eqnarray}
with $m=w,w_c$ and $n=d,e$. 
\begin{eqnarray}
N^{t_{1}t_{3}}_{\omega\omega,pqr}(\br_1,\br_{1'})&=&
\Big[X^{t_{1}t_{3}}_{\omega d,q}(\br_1,\br_2)
\xi^{k_2k_3k_1}_{121'}C_{d,qr}^{t_{3}}(\br_2)|
X_{d\omega,r}^{t_{3}t_{1}}(\br_2,\br_{1'})+
N_{d\omega,r}^{t_{3}t_{1}}(\br_2,\br_{1'})\Big]+\\
\nonumber
&&\Big[X^{t_{1}t_{3}}_{\omega e,q}(\br_1,\br_2)
\xi^{k_2k_3k_1}_{121'}C_{e,qr}^{t_{3}}(\br_2)|
X_{d\omega,r}^{t_{3}t_{1}}(\br_2,\br_{1'})+
N_{d\omega,r}^{t_{3}t_{1}}(\br_2,\br_{1'})\Big]+\\
\nonumber
&&\Big[X^{t_{1}t_{3}}_{\omega d,q}(\br_1,\br_2)
\xi^{k_2k_3k_1}_{121'}C_{e,qr}^{t_{3}}(\br_2)|
X_{e\omega,sr}^{t_{3}t_{1}}(\br_2,\br_{1'})+
N_{e\omega,r}^{t_{3}t_{1}}(\br_2,\br_{1'})\Big]
\,\,,
\\
N^{t_{1}t_{2}t_{3}}_{\omega n,pqr}(\br_1,\br_{2})&=&
\Big[X^{t_{1}t_{3}}_{\omega d,q}(\br_1,\br_3)
\xi^{k_2k_3k_1}_{132}C_{d,qr}^{t_{3}}(\br_3)|
X_{dn,r}^{t_{3}t_{2}}(\br_3,\br_{2})+
N_{dn,r}^{t_{3}t_{2}}(\br_3,\br_{2})\Big]+\\
\nonumber
&&\Big[X^{t_{1}t_{3}}_{\omega e,q}(\br_1,\br_3)
\xi^{k_2k_3k_1}_{132}C_{e,qr}^{t_{3}}(\br_3)|
X_{dn,r}^{t_{3}t_{2}}(\br_3,\br_{2})+
N_{dn,r}^{t_{3}t_{2}}(\br_3,\br_{2})\Big]+\\
\nonumber
&&\Big[X^{t_{1}t_{3}}_{\omega d,q}(\br_1,\br_3)
\xi^{k_2k_3k_1}_{132}C_{e,qr}^{t_{3}}(\br_3)|
X_{en,r}^{t_{3}t_{2}}(\br_3,\br_{2})+
N_{en,r}^{t_{3}t_{2}}(\br_3,\br_{2})\Big]\\
\,\,
\end{eqnarray}
also in the above equations we used $n=d,e$. 
In the following equations we have that $m,n=c,w_c$.
\begin{eqnarray}
\nonumber
N^{t_{1}t_{3}}_{mnx,pqr}(1,{2})&=&
\Big[X^{t_{1}}_{mc,q}(1,3)
\xi^{k_2k_3k_1}_{132}\frac{\Delta^{k_3}}{2}
C_{e,qr}^{t_{3}}(3)|
X_{cn}^{t_{3}}(3,{2})+N_{cnx}^{t_{3}}(3,{2})+
N_{\rho n}^{t_{3}}(3,{2}) \Big]+\\
&&(1-\delta_{r,1}) \times \\
\nonumber & &
\Big[X^{t_{1}}_{mc}(1,3)
\xi^{k_2k_3k_1}_{132}\frac{\Delta^{k_2}}{2}
C_{e,qr}^{t_{3}}(3)|
X_{cn,r}^{t_{3}}(3,{2})+
N_{cnx,r}^{t_{3}}(3,{2})+
N_{\rho n,r}^{t_{3}}(3,{2}) \Big]-\\
\nonumber
&& \hspace*{-3mm}
\delta_{k_1 1}\delta_{k_2 1}\delta_{k_3 1} \Big\{
\Big[X^{t_{1}}_{mcj,q}(1,3)
\frac{1}{2}
C_{e,qr}^{t_{3}}(3) |
X_{cnj}^{t_{3}}(3,{2})+
N_{cnxj}^{t_{3}}(3,{2})+
N_{\rho nj}^{t_{3}}(3,{2})\Big]\\
\nonumber
&&\hspace*{-3mm}
+(1-\delta_{r,1})\Big[X^{t_{1}}_{mcj}(1,3)
\frac{1}{2}
C_{e,qr}^{t_{3}}(3) |
X_{cnj,r}^{t_{3}}(3,{2})+
N_{cnxj,r}^{t_{3}}(3,{2})+
N_{\rho nj,r}^{t_{3}}(3,{2})\Big] \Big\} 
\,\,,
\\
\nonumber
N^{t_{1}t_{3}}_{m\rho,pqr}(1,{2})&=&
\Big[X^{t_{1}}_{mc,q}(1,3)
\xi^{k_2k_3k_1}_{132}\frac{\Delta^{k_3}}{2}
C_{e,qr}^{t_{3}}(3)|
-\rho_{o}^{t_{3}}(3,{2})+
N_{c\rho}^{t_{3}}(3,{2})+
N_{\rho}^{t_{3}}(3,2)\Big]\\
\nonumber
&&+(1-\delta_{r,1})
\Big[X^{t_{1}}_{mc}(1,3)
\xi^{k_2k_3k_1}_{132}\frac{\Delta^{k_2}}{2}
C_{e,qr}^{t_{3}}(3)|
N_{c\rho,r}^{t_{3}}(3,{2})+
N_{\rho,r}^{t_{3}}(3,2)\Big]-\\
\nonumber
&&\delta_{k_1 1}\delta_{k_2 1}\delta_{k_3 1} \Big\{
\Big[X^{t_{1}}_{mcj,q}(1,3)
\frac{1}{2}
C_{e,qr}^{t_{3}}(3) |
-\rho_{oj}^{t_{3}}(3,{2})+
N_{c\rho j}^{t_{3}}(3,{2})+
N_{\rho j}^{t_{3}}(3,2)\Big] \\
&&+(1-\delta_{r,1})\Big[X^{t_{1}}_{mcj}(1,3)
\frac{1}{2}
C_{e,qr}^{t_{3}}(3)|
N_{c\rho j,r}^{t_{3}}(3,{2})+
N_{\rho j,r}^{t_{3}}(3,2)\Big] \Big\} 
\,\,,
\\
N^{t_{1}t_{3}}_{\rho,pqr}(1,{2})&=&
-\Big[\rho^{t_{1}}_{0}(1,3)
\xi^{k_2k_3k_1}_{132}\frac{\Delta^{k_2}}{2}
C_{e,qr}^{t_{3}}(3)|
N_{c\rho,r}^{t_{3}}(3,{2})\Big]\\ \nonumber & &
-\Big[\rho^{t_{1}}_{0}(1,3)
\xi^{k_2k_3k_1}_{132}\frac{\Delta^{k_2}}{2}
(C_{e,qr}^{t_{3}}(3)-1)|
N_{\rho,r}^{t_{3}}(3,{2})-\delta_{r,1}\rho^{t_{3}}_{0}(3,2)\Big]\\
\nonumber
&&+\delta_{k_1 1}\delta_{k_2 1}\delta_{k_3 1} \Big\{
\Big[\rho^{t_{1}}_{0j}(1,3)
\frac{1}{2}
C_{e,qr}^{t_{3}}(3)|
N_{c\rho j,r}^{t_{3}}(3,{2})\Big]\\ \nonumber & &
+\Big[\rho^{t_{1}}_{0j}(1,3)
\frac{1}{2}
(C_{e,qr}^{t_{3}}(3)-1)|
N_{\rho j,r}^{t_{3}}(3,{2})-
\delta_{r,1}\rho^{t_{3}}_{0j}(3,2)\Big] \Big\} 
\,\,,
\\
\nonumber
N^{t_{1}t_{3}}_{mnxj,pqr}(1,{2})&=&
\delta_{k_1 1}\delta_{k_2 1}\delta_{k_3 1} 
\Big\{ \Big[X^{t_{1}}_{mc,q}(1,3)
\frac{1}{2}
C_{e,qr}^{t_{3}}(3)|
X_{cnj}^{t_{3}}(3,{2})+
N_{cnxj}^{t_{3}}(3,{2})+
N_{\rho nj}^{t_{3}}(3,2)\Big]+\\
\nonumber
&&(1-\delta_{r,1}) \Big[X^{t_{1}}_{mc}(1,3)
\frac{1}{2}
C_{e,qr}^{t_{3}}(3)|
X_{cnj,r}^{t_{3}}(3,{2})+
N_{cnxj,r}^{t_{3}}(3,{2})+
N_{\rho nj,r}^{t_{3}}(3,2)\Big]+\\
\nonumber
&&
\Big[X^{t_{1}}_{mcj,q}(1,3)
\frac{1}{2}
C_{e,qr}^{t_{3}}(3)|
X_{cn}^{t_{3}}(3,{2})+
N_{cnx}^{t_{3}}(3,{2})+
N_{\rho n}^{t_{3}}(3,2)\Big]+\\
\nonumber
&&(1-\delta_{r,1}) \Big[X^{t_{1}}_{mcj}(1,3)
\frac{1}{2}
C_{e,qr}^{t_{3}}(3)|
X_{cn,r}^{t_{3}}(3,{2})+
N_{cnx,r}^{t_{3}}(3,{2})+
N_{\rho n,r}^{t_{3}}(3,2)\Big] \Big\} 
\,\,,
\\ \\
\nonumber
N^{t_{1}t_{3}}_{m \rho j,pqr}(1,{2})&=&
\delta_{k_1 1}\delta_{k_2 1}\delta_{k_3 1} 
\Big\{\Big[X^{t_{1}}_{mc,q}(1,3)
\frac{1}{2}
C_{e,qr}^{t_{3}}(3)|
-\rho_{oj}^{t_{3}}(3,{2})+
N_{c\rho j}^{t_{3}}(3,{2})+
N_{\rho j}^{t_{3}}(3,2)\Big]\\
\nonumber
&&+(1-\delta_{r,1})
\Big[X^{t_{1}}_{mc}(1,3)
\frac{1}{2}
C_{e,qr}^{t_{3}}(3)|
N_{c\rho j,r}^{t_{3}}(3,{2})+
N_{\rho j,r}^{t_{3}}(3,2)\Big]\\
\nonumber
&&+\Big[X^{t_{1}}_{mcj,q}(1,3)
\frac{1}{2}
C_{e,qr}^{t_{3}}(3)|
-\rho_{o}^{t_{3}}(3,{2})+
N_{c\rho }^{t_{3}}(3,{2})+
N_{\rho }^{t_{3}}(3,2)\Big]+\\
%\nonumber
&&(1-\delta_{r,1})
\Big[X^{t_{1}}_{mcj}(1,3)
\frac{1}{2}
C_{e,qr}^{t_{3}}(3)|
N_{c\rho,r}^{t_{3}}(3,{2})+
N_{\rho ,r}^{t_{3}}(3,2)\Big]\Big\} 
\,\,,
\\
\nonumber
N^{t_{1}t_{3}}_{\rho j,pqr}(1,{2})&=&
-\delta_{k_1 1}\delta_{k_2 1}\delta_{k_3 1} \Big\{
\Big[\rho^{t_{1}}_{0}(1,3)
\frac{1}{2}
C_{e,qr}^{t_{3}}(3)|
N_{c\rho j,r}^{t_{3}}(3,{2})\Big]\\ \nonumber & &
+\Big[\rho^{t_{1}}_{0}(1,3)
\frac{1}{2}
(C_{e,qr}^{t_{3}}(3)-1)|
N_{\rho j,r}^{t_{3}}(3,{2})-\delta_{r,1}\rho^{t_{3}}_{0j}(3,2)\Big]\\
\nonumber
&&+
\Big[\rho^{t_{1}}_{0j}(1,3)
\frac{1}{2}
C_{e,qr}^{t_{3}}(3)|
N_{c\rho ,r}^{t_{3}}(3,{2})\Big]\\ 
%\nonumber 
&~&
+\Big[\rho^{t_{1}}_{0j}(1,3)
\frac{1}{2}
(C_{e,qr}^{t_{3}}(3)-1)|
N_{\rho ,r}^{t_{3}}(3,{2})-\delta_{r,1}\rho^{t_{3}}_{0}(3,2)\Big] \Big\}
\,\,.  
\end{eqnarray}

The values of the $\xi^{k_1,k_2,k_3}_{ijk}$ coefficients are given in
Ref. \cite{pan79}. 
%
%
% THE BIBLIOGRAPHY
%
%\bibliography{altro,miei,fhnc,em,nuovo}

%
%
%
% Table
%
\clearpage
\newpage 
%
% Occupation numbers
%
%%%%
%
% FIGURE
%
% END
%
\clearpage
\newpage
%
%
% END
%
\end{document}